\documentclass[a4paper,12pt]{article}

\usepackage{amsmath}
\usepackage{amssymb}
\usepackage{amsfonts}

\usepackage{wasysym}

\usepackage{multirow}
\usepackage{tabularx}
\usepackage{adjustbox}

\usepackage{import}
\usepackage{tikz, tikz-3dplot}
\usetikzlibrary{shapes,arrows.meta,chains,shapes.multipart,spy,backgrounds}
\usepackage{pgfplots}
\usepgfplotslibrary{fillbetween} 
\pgfplotsset{compat=1.18}
\usepackage{pgfplotstable}
\usepackage{xcolor}
\usepackage{overpic}
\usepackage{subcaption}
\usepackage{float}

\usepackage{enumitem}

\usepackage{comment}

\usepackage{doi}
\usepackage{url}
\usepackage[numbers,sort&compress]{natbib}
\usepackage{authblk}

\newcommand{\ksi}{\xi}

\newcommand{\GF}[1]{\mathcal{#1}}

\newcommand{\tikzcircle}[2][red,fill=red]{\tikz[baseline=-0.5ex]\draw[#1,radius=#2] (0,0) circle ;}%

\renewcommand{\Re}{\ensuremath{\text{Re}}}
\newcommand{\Ga}{\ensuremath{\text{Ga}}}
\def\gz #1{\mbox{\boldmath $\mit #1$}}

\tikzset{spy scope/.append code={
   \patchcmd\tikz@atend@scope{\egroup\egroup}{%
      \pgf@remember@layerlist@globally%
      \egroup\egroup%
      \pgf@restore@layerlist@from@global%
   }{}{}
}}

\title{\vspace{-2em} 
Fluctuations, Clustering, and Interaction-Driven Dynamics in Sedimenting Particles at Low Galileo Numbers: A Neural Network Approach
\\[1ex] 
\large\textit{Submitted to the Journal of Fluid Mechanics (under review)}}

\author[1]{Nejc Vovk\thanks{Corresponding author: nejc.vovk@um.si}}
\author[2]{Jana Wedel}
\author[2,3]{Paul Steinmann}
\author[1]{Jure Ravnik}

\affil[1]{Faculty of Mechanical Engineering, University of Maribor, Slovenia}
\affil[2]{Institute of Applied Mechanics, University of Erlangen-Nürnberg, Germany}
\affil[3]{Glasgow Computational Engineering Center, University of Glasgow, Scotland}

\date{}

\begin{document}
\maketitle

\begin{abstract}
In this study, we investigate the behaviour of sedimenting solid particles and the influence of microscopic particle dynamics on the collective motion of a sedimenting cloud. Departing from conventional direct numerical simulations (DNS), we introduce a novel machine learning framework, the Interaction-Decomposed Neural Network (IDNN), to model hydrodynamic particle interactions. The IDNN acts as a black-box module within a Lagrangian solver, predicting the particle drag force based on the relative positions of the nearest neighbours. This enables the recovery of force fluctuations, capturing effects previously accessible only through DNS. Our results show an increase in collective settling velocity in the dilute regime, consistent with earlier experimental and numerical studies, which we attribute to (i) fluctuations in the streamwise particle force around a value that is lower than the Stokes limit and (ii) the formation of particle clusters sedimenting at enhanced velocities. These fluctuations originate from persistent entrainment and ejection of particles in and out of the long, diffusive wakes generated by upstream particles at low Galileo numbers. Energy spectra of particle velocity fluctuations reveal a scale-dependent transfer of fluctuation energy, analogous to a turbulent-like cascade, with pronounced large-scale fluctuations at higher volume fractions. At very low volume fractions, fluctuation intensity and energy spectrum amplitudes diminish, though hydrodynamic interactions still remain appreciable.

\end{abstract}


\section{Introduction}
The sedimentation of solid particles in a quiescent medium is a fundamental process relevant to atmospheric science, environmental engineering, and industrial applications such as aerosol drug delivery and particulate pollution control \citep{clarkTransportModelingEnvironmental1996}. The Galileo number $\Ga$ plays a role in particle sedimentation, analogous to that of the particle Reynolds number $\Re_{\text{p}}$. Both nondimensionalize inertial versus viscous forces in the flow. However, by construction $\Re_{\text{p}}=|\mathbf{u}_{\text{rel}}|d_\text{p}/\nu$ uses a specified velocity scale $\mathbf{u}_{\text{rel}}$ (for example an imposed relative flow or measured relative particle velocity), whereas the Galileo number is defined using buoyancy (gravity) as the driving scale. In practice $\Ga$ is set by particle and fluid properties (density ratio, size, gravity, viscosity) and thus embodies the buoyancy--based inertial scale, without needing a prescribed velocity. In this way $\Ga$ serves as a convenient control parameter for sedimentation flows -- for example, low $\Ga$ corresponds to Stokes (viscous--dominated) sedimentation, while higher $\Ga$ implies increasing inertial effects and higher $\Re$. 

Even in non--turbulent situations, where a single particle settles in a quiescent fluid, the trajectory of the sedimenting particle can be unstable i.e. showing non--negligible lateral velocity fluctuations \citep{jennyInstabilitiesTransitionSphere2004}. The emergence of particle trajectory instabilities during sedimentation is attributed to the formation of wakes around the particle. These instabilities arise when gravity-driven inertial forces begin to dominate over viscous forces, a phenomenon that parallels classical $\Re$ number--based flow regime transitions. The problem has been studied for sedimentation of single spheres \citep{cabrera-boomanPathInstabilitiesDrag2024}, multiple spheres and suspensions \citep{machesSettlingTwoRigidly2024, uhlmannSedimentationDiluteSuspension2014}, as well as non--spherical particles \citep{gaiDynamicsWakesFreely2024,moricheClusteringLowaspectratioOblate2023}. In parallel with Reynolds number-based analyses, the Galileo number has been employed to characterize sedimentation regimes. High $\Ga$ values are indicative of increasingly unstable and chaotic particle trajectories during sedimentation \citep{cabrera-boomanPathInstabilitiesDrag2024}. In contrast, at low $\Ga$ numbers, where viscous forces dominate over gravitational and inertial effects, the sedimenting behaviour of a single particle exhibits a steady, vertical trajectory, with an axisymmetric wake, similar to the wake behind a fixed sphere at very low $\Re$ numbers \citep{raaghavPathInstabilitiesFreely2022}. Here, the dominant viscous forces dictate the fluid momentum transport via diffusion. In the fundamental studies by \citet{batchelorSedimentationDiluteDispersion1972, batchelorIntroductionFluidDynamics2000}, it has been shown that, in this regime, the disturbance caused by a sphere moving through a fluid extends significantly into the surrounding medium, with the velocity decaying radially as $1/r$ from the sphere. Batchelor also argued that such velocity field disturbance can have a significant influence on other spheres sedimenting nearby, even many diameters apart. 

The assumptions associated with such viscous-dominated flow regime allowed \citet{stokesEffectInternalFriction1851} to analytically solve the governing Navier-Stokes equations around a single spherical particle, to obtain the well known drag model \citep{stokesEffectInternalFriction1851},
\begin{equation}\label{eq:stokesDrag}
    |\mathbf{F}|_{\text{d, Stokes}} = 3 \pi \mu_\text{f} d_\text{p} | \mathbf{u}_{\text{rel}} |,
\end{equation}
where $\mu_\text{f}$ is the dynamic viscosity of the carrier fluid, $d_\text{p}$ the particle diameter, and $| \mathbf{u}_{\text{rel}} |$ the particle-fluid relative velocity magnitude. In a system comprising numerous small particles sedimenting in a fluid under Stokes flow conditions, long-range hydrodynamic interactions play a critical role in shaping both individual trajectories and collective dynamics. \citep{janosiChaoticParticleDynamics1997,pineChaosThresholdIrreversibility2005, metzgerIrreversibilityChaosRole2010}. The term "hydrodynamic interaction" refers to the fluid-mediated forces that occur between particles in the suspension due to their movement in the fluid. Specifically, when relative velocity exists between the particle and the fluid, a disturbance in the surrounding fluid takes place, creating flow fields that influence neighbouring particles. These flow fields can cause the particles to reorient, move, or cluster, thereby affecting the collective behaviour of the suspension. Naturally, the long-range nature stems from the prevalent viscous forces, that exhibit diffusive transport of momentum from the particle onto the surrounding fluid.

Hydrodynamic interactions are challenging to model due to their multibody character, and the need to account for complex flow reflections between particles \citep{langSoftMatterAqueous2016}. In case of a real multiphase system of many particles, this multibody type of problem demonstrates conceptual similarities to the type of problem principally studied with celestial mechanics in astronomy, where the pioneering work has been done by \citet{poincareThreebodyProblemEquations1890}, who described a multibody system as chaotic. If we consider the particle hydrodynamic interaction as a multibody problem, the movement of the individual particles could as well be described as intrinsically chaotic. The movement prediction of both the celestial bodies and the particles in fluid requires analysing the collective behaviour of multiple objects under specific forces. In celestial mechanics, the problem focuses on predicting the motion of individual celestial bodies through gravity, which becomes increasingly more complex and non--linear as the number of bodies increases \citep{pealeCelestialMechanics2023}. Similarly, in fluid dynamics, understanding the motion of multiple particles moving through a fluid requires accounting for hydrodynamic forces among them. Many researchers have previously studied this problem from a multibody perspective by making use of the so--called grand resistance matrix. Readers interested in foundational work on this topic are encouraged to refer to some interesting earlier studies, such as: \citet{mazurManysphereHydrodynamicInteractions1982, ganatosNumericalsolutionTechniqueThreedimensional1978, durlofskyDynamicSimulationHydrodynamically1987}.

All the aforementioned arguments suggest that suspended microparticles, accompanied by a low $\Ga$ number, do not sediment along steady, vertical trajectories. Instead, such particles may exhibit chaotic motion driven by hydrodynamic interactions. A key question, therefore, is under what conditions the hydrodynamic interactions become strong enough that their full, multibody complexity must be considered? According to \citet{corsonHydrodynamicInteractionsAerosol2018}, these interactions become significant when the particle Knudsen number—defined by the ratio of particle size to the mean free path of fluid molecules—is sufficiently large. This ensures that the continuum hypothesis remains valid, allowing for meaningful hydrodynamic interaction modelling. The mechanism behind hydrodynamic interaction and its effect on collective cloud behaviour is also different, whether large particles in dense suspensions or small particles in a dilute regime are considered. Numerous studies have shown that in the first regime, these interactions lead to hindered settling velocities \citep{liHinderedSettlingLognormally2024, yaoEffectsParticleClustering2021}. For example, \citet{penlouExperimentalMeasurementEnhanced2023} demonstrated that large particles ($467~\mu\text{m}$) sedimenting in air experience a reduction in terminal velocity, while smaller particles ($78~\mu\text{m}$) display an increase in mean terminal velocity with increasing volume fraction. Both effects on the terminal settling velocity are attributed to the so--called clustering of particles. For larger particles, the hindered settling velocity is said to be a consequence of the inertial effects inside clusters, such as inertial wake formation \citep{zaidiHinderedSettlingVelocity2015, yaoEffectsParticleClustering2021} and the so-called drafting--kissing--tumbling effect \citep{fortesNonlinearMechanicsFluidization1987,hamHinderedSettlingHydrodynamic1988, yaoEffectsParticleClustering2021}. Another interesting matter is the analysis of the particle clustering itself. Spherical particle clustering in dilute suspensions is said to be more noticeable as $\Ga$ number increases \citep{uhlmannSedimentationDiluteSuspension2014}.

While these mechanisms of the hindered settling velocity of large particles in dense suspensions are generally agreed upon, the long-range hydrodynamic interaction of small particles, and their behaviour in dilute suspensions is still a matter of ongoing debate. On top of that, the microscopic dynamics of particles and its effect on the macroscopic behaviour of the suspension during sedimentation is still largely uncategorized. This can be explained by the multitude of interacting physical parameters, such as the particle-fluid density ratio, particle size, fluid viscosity, particle volume fraction etc., as discussed in a review by \citet{brandtParticleLadenTurbulenceProgress2022}. Nonetheless, a comprehensive analysis of particle sedimenting across the full spectrum of interacting parameters should be addressed in future research.

Recently, two primary approaches have been developed to compute force fluctuations on a spherical particle in the presence of surrounding particles. \citet{akikiPairwiseInteractionExtended2017,akikiPairwiseinteractionExtendedPointparticle2017} presented the so-called pairwise interaction extended point-particle approach (PIEP), that assumes that neighbouring particles interact with the reference particle only in pairs. \citet{seyed-ahmadiPhysicsinspiredArchitectureNeural2022} integrated this approach into an ANN architecture. They predicted the streamwise force on a particle based on the positions of the neighbouring particles, which they input into the ANN one at a time instead of all at once, thereby capturing the pairwise interactions. The reason for implementing the PIEP into an ANN architecture was based on the findings by previous authors \citep{heSupervisedMachineLearning2019, balachandarParticleresolvedAccuracyEuler2020} who tackled this task by creating a classic dense ANN architecture, inputting the neighbouring coordinates all at once. The subpar performance of such architectures was attributed to the well-known "curse of dimensionality," which refers to the exponential increase in computational complexity and data requirements as the number of input dimensions grows, making effective learning and generalization increasingly difficult. 

The second approach was developed by \citet{seyed-ahmadiMicrostructureinformedProbabilitydrivenPointparticle2020}, who called it the microstructure-informed probability-driven point-particle (MPP) approach, which relies on a probability-driven regression for computing the correction to the force on a particle. Both the PIEP and the MPP rely to some extent on empirical modelling. The PIEP incorporates the so-called undisturbed velocity paradigm, in which interparticle forces are calculated using the fluid velocity that would exist in the absence of the particle's own disturbance. 


While the undisturbed velocity approach also captures multibody interactions, the goal of our model is to achieve this without requiring access to the undisturbed velocity field. Instead, we want it to rely solely on the local particle arrangement to capture multibody behaviour and helps us better understand its effect on sedimentation as a whole. Additionally, we extend the pairwise-interaction assumption in order to more physically capture the behaviour through higher order interactions. The paper is divided into two main parts; in the first part, we present and discuss the results of numerous simulations, needed to construct our database. Next, we introduce the novel interaction-decomposed neural network (IDNN) architecture and discuss all the past findings contributing to its structure. In the second part, we present the results of mineral dust sedimentation simulations as well as their comparison with some other authors. At the end, we discuss the findings of how hydrodynamic interactions at low $\Ga$ numbers influence the collective cloud sedimentation from various perspectives.

\textbf{Notation:} In this paper, we express physical vectors using lowercase bold letters, e.g. $\mathbf{a}$ and higher order tensors using uppercase bold letters, e.g. $\mathbf{A}$. The individual tensor coefficients can be assembled in a coefficient matrix, which we will denote with underlined lowercase letters for vectors, $\underline{a}$ and underlined uppercase letters for higher dimensional tensors, $\underline{A}$. When performing a transformation of a coordinate system, the tensor coefficients change 
\begin{equation*}
    \underline{a}' = \underline{R} \, \underline{a}, \hspace{0.5cm} \underline{A}' = \underline{R} \, \underline{A} \, \underline{R}^\text{T},
\end{equation*}
where $\underline{R}$ is the rotation matrix. Throughout this work, we adopt this notation consistently. Any exceptions will be explicitly noted. For example, although the force $\mathbf{F}$ is formally a first-order tensor, we retain uppercase vector notation $\mathbf{F}$ in accordance with common convention in physics and engineering literature.
\section{Dataset construction}
\subsection{Overview}
\begin{figure}[h]
    \centering
    \begin{tikzpicture}
        \node{\includegraphics[width=\textwidth, trim={0cm 6cm 0cm 4cm}, clip]{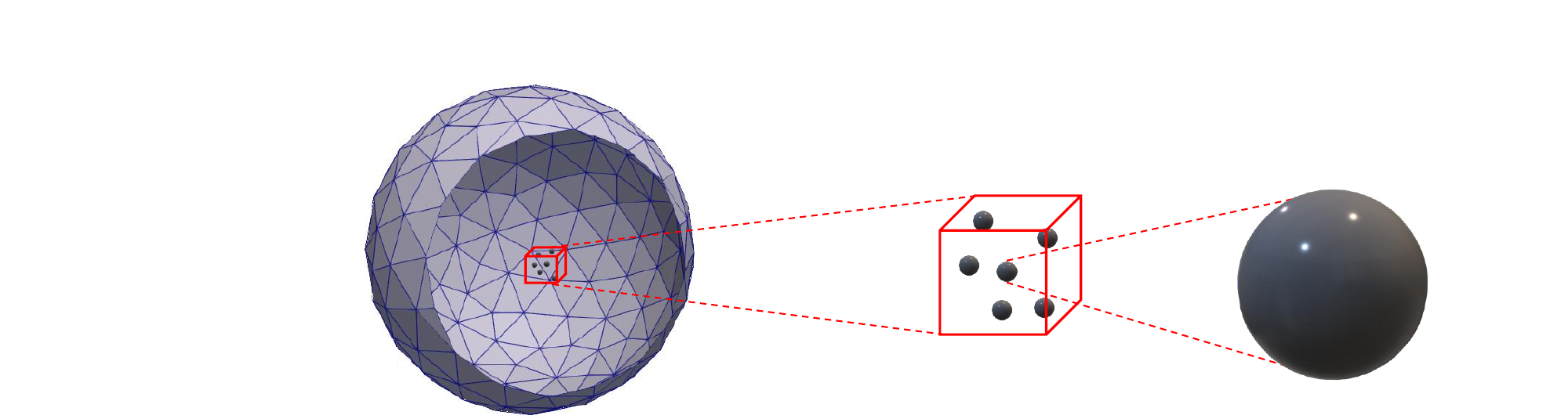}};
        \draw[-Stealth] (-3.8,-1.8) -- ++ (0.5,0.5);
        \draw[-Stealth] (-0.7,1.3) -- ++ (-0.5,-0.5);
        \node at (0.8,1.6) {$d_{\text{domain}} = 1024 d_{\text{cl}}$};

        \draw[-Stealth] (3.1,0.75) -- ++ (-0.5,-0.5);
        \draw[-Stealth] (0.8,-1.75) -- ++ (0.5,0.5);
        \node at (0.8,-2) {$d_{\text{cl}}$};

        \draw[-Stealth] (3.6,-1.8) -- ++ (0.5,0.5);
        \draw[-Stealth] (5.8,0.5) -- ++ (-0.5,-0.5);
        \node at (3.9,-2) {$d_\text{p}$};

        \node at (-6,0.8) {$\mathbf{u}_\text{f}$};
        \draw[gray,->] (-6,0.4) -- ++ (0.5,0);
        \draw[gray,->] (-6,0.2) -- ++ (0.5,0);
        \draw[gray,->] (-6,0) -- ++ (0.5,0);
        \draw[gray,->] (-6,-0.2) -- ++ (0.5,0);
        \draw[gray,->] (-6,-0.4) -- ++ (0.5,0);
        \draw[gray,->] (-6,-0.6) -- ++ (0.5,0);
        \draw[gray,->] (-6,-0.8) -- ++ (0.5,0);
        \draw[gray,->] (-6,-1) -- ++ (0.5,0);

        \node at (-5.8,-2.2) {vertical};
        \node at (-4.7,-2.5) {spanwise};
        \node at (-4.3,-3) {streamwise};
        \draw[->] (-6,-3) -- ++ (0.5,0);
        \draw[->] (-6,-3) -- ++ (0,0.5);
        \draw[->] (-6,-3) -- ++ (0.3535,0.3535);
    \end{tikzpicture}
    \caption{The domain is spherical, with a reference particle, for which the drag force is measured, at the centre. A cluster of five other randomly arranged particles is inserted around the reference particle. We measure the force on the reference particle in streamwise, spanwise and vertical directions, while changing the positions of the five neighbouring particles. By varying the radial distance of the neighbouring particles, we simulate the volume fraction range $10^{-8} < \varphi < 10^{-3}$. The domain boundary as well as the particle surface are discretized.}
    \label{fig:problemDefinition}
\end{figure}
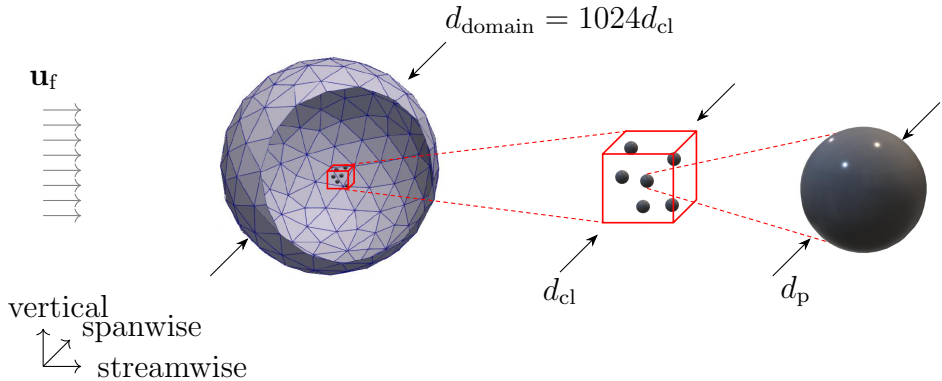

We performed a total of $14~000$ simulations with rigid spherical particles subjected to a uniform plug flow. The domain is spherical, with a reference particle for which the drag force is measured at the centre. A cluster of five other randomly arranged particles is inserted around the reference particle. All particles are spherical with diameter $d_\text{p}$. The outer boundary, for which the flow boundary conditions are set, is $1024$ times larger than the particle cluster, as shown in Fig. \ref{fig:problemDefinition}. Due to the diffusive nature of momentum transfer in Stokes flow, such a large size difference is necessary to ensure that the boundary conditions do not affect the force measurement \citep{straklNumericalDragLift2022,straklModelTranslationRotation2022,anderssonForcesTorquesProlate2019}. For solving the flow around the cluster of particles, the boundary element method (BEM) is employed, using the in-house code Andromeda \citep{ravnikAnalyticalExpressionsSingular2023,ravnikAndromedaBEMBased2025}. We measure the force on the reference particle in streamwise, spanwise and vertical directions, while changing the positions of the five neighbouring particles. By varying the radial distance of the neighbouring particles, we simulate the volume fraction range $10^{-8} < \varphi < 10^{-3}$.

It is important to acknowledge that the decision to include only the five nearest neighbours in the calculation is not based on a specific physical assumption, but rather adopted as a modelling simplification. The five-neighbour assumption is based the following:
\begin{itemize}
    \item \citet{seyed-ahmadiMicrostructureinformedProbabilitydrivenPointparticle2020} concluded, that the inclusion of $5$--$10$ neighbours is sufficient for lateral force prediction, while up to $20$ neighbours is needed for prediction of streamwise force. In a later study \citet{seyed-ahmadiPhysicsinspiredArchitectureNeural2022} built a pairwise-interaction ANN and concluded that including more than $5$ neighbours into training only marginally improves the ANN performance. They also found that, observing the ANN weight values, the strongest weights correspond to the closest neighbours. The latter is also later proved in our own efforts, by plotting the correlations in Fig. \ref{fig:correlation}.
    \item The dataset is constructed by the DNS simulation of Stokes flow around particles using BEM. As already mentioned, to obtain a BEM solution, only the geometry surfaces must be discretized. The underlying procedure involves obtaining a solution of the system of equations that results in a dense system matrix, as opposed to the FVM or FEM, which results in a sparse matrix \citep{ravnikAnalyticalExpressionsSingular2023}. Due to this fact, regular sparse system solvers could not be used to obtain a solution, and the operation complexity is of order $\mathcal{O}(n^2)$ \citep{wrobelBoundaryElementMethod2002}. The inclusion of every additional particle significantly increases the computational complexity, especially on the long run, when performing numerous simulations. Nonetheless, in our case, BEM enabled us to obtain solutions significantly faster than the FVM.
\end{itemize}

\subsection{DNS of Stokes flow}
Here, a brief breakdown of the governing equations as well as the method for obtaining a solution using BEM will be outlined. The more detailed procedure of particle force computation is given in Appendix \ref{appA}.

We consider the steady incompressible flow of a Newtonian fluid at very small Reynolds numbers, i.e. $\Re\ll1$, where we can neglect the advection term in the Navier-Stokes equations, leading to the equations of Stokes flow:
\begin{equation}
	\mathbf{\nabla}\cdot\mathbf{u}_\text{f} = 0, \qquad \mathbf{\nabla}\cdot \gz \sigma+\rho_\text{f} \mathbf{g} = 0.
\end{equation}
Here $\mathbf{u}_\text{f}$ is the flow velocity, $\rho_\text{f}$ is the fluid density and $\mathbf{g}$ is the gravitational acceleration. The Cauchy stress tensor $\mathbf{\sigma}$ is defined as 
$	\gz \sigma = -p\mathbf{I}+ \gz \tau,$
where $p$ is the pressure, $\mathbf{I}$ the identity tensor, and $\gz \tau$ the viscous stress tensor. A Newtonian model for the viscous stress tensor 
leads to the following form of the Stokes equation
\begin{equation}
	-\mathbf{\nabla} p + \mu_\text{f} \nabla^2 \mathbf{u}_\text{f} +\rho_\text{f} \mathbf{g} = 0,
\end{equation}
where $\mu_\text{f}$ is the fluid viscosity. Finally, we recognize that gravity is a conservative force, which may be written as a gradient of the gravitational potential and introduce the modified pressure as $p^*=p-\rho_\text{f} \Phi$, where $\mathbf{g} = \mathbf{\nabla}\Phi$. With this, the final form of the Stokes equation reads
\begin{equation}\label{eq:StokesEquation}
	-\mathbf{\nabla} p^* + \mu_\text{f} \nabla^2 \mathbf{u}_\text{f}  = 0.
\end{equation}

The advocated BEM for this problem possesses two advantages over traditional volume based methods: first, only the particles and the outer domain need to be discretized (not the whole very large domain volume), and second, the traction values are a part of the solution procedure (no need to numerically calculate derivatives in post-processing, which is prone to discretization errors). Having boundary traction $\mathbf{q}$ readily available for each boundary element at the particle surface makes calculation of the drag force simple:
\begin{equation}
	\mathbf{F} = \int_\Gamma\mathbf{q} \text{d}\Gamma = 
	\sum_l \mathbf{q}^{(l)}A_l 
\end{equation} 
The nondimensional version of Stokes drag force in Eq. (\ref{eq:stokesDrag}) can be expressed as the drag coefficient $c_\text{D}=|\mathbf{F}|_{\text{d, Stokes}}/[A\frac{1}{2}\rho_\text{f} |\mathbf{u}_\text{f} - \mathbf{u}_\text{p}|^2]=24/\text{Re}_\text{p}$. The results and models in this paper are presented and developed in a non-dimensional fashion, expressing force as the product of Reynolds number and drag coefficient, i.e. $\text{Re}_\text{p} c_\text{D}$. In this description, the Stokes drag for a single particle in plug flow takes the value of $\text{Re}_\text{p} c_\text{D}=24$.

Since the objective of this study is to investigate multibody particle interactions using a trained ANN model—based on a training dataset composed of numerous BEM simulations—it is essential to select a computational mesh that ensures sufficient accuracy while also maintaining reasonable computational efficiency to enable a large number of simulations. The procedure for choosing the adequate mesh is shown in Appendix \ref{appC}.

\subsection{Interpretation of the BEM results}
\begin{figure}[!h]
    \includegraphics{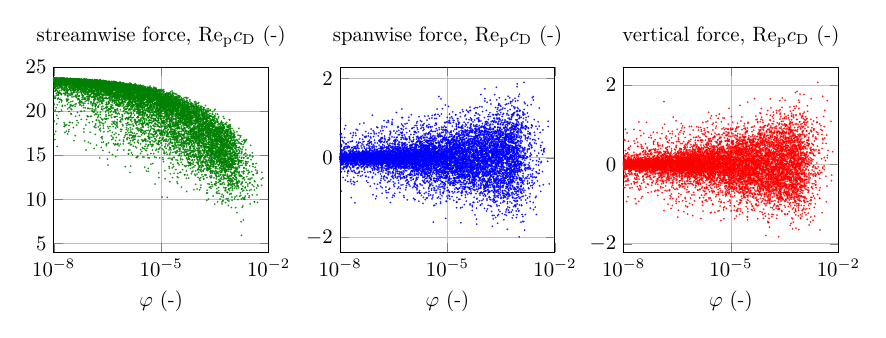}
    \caption{BEM simulation results for all three force components. For convenience, only $\sim 1000$ data points are shown in plots.}
        \label{fig:BEMresults}
\end{figure}

For each BEM simulation, we estimated the volume fraction based on the fact that for Poisson-distributed set of points in 3D, the expected distance $d_k$ to the $k$-th nearest neighbour scales as, \citep{clarkDistanceNearestNeighbor1954}
\begin{equation}
    d_k \approx \left[\frac{k V_p}{\alpha \varphi}\right]^{1/3}
\end{equation}
where $\alpha$ is a constant that depends on the spatial dimension. In 3D, it is approximately $\alpha \approx 4\pi/3$. Accordingly, the volume fraction for the $i$-th BEM simulation can be determined based on the distance to the fifth neighbour as
\begin{equation}\label{eq:volFrac_definition}
    \varphi_i = \frac{15 V_\text{p}}{4 \pi d_5^3},
\end{equation}
where $V_\text{p}$ is the particle volume and $d_5$ the distance to the fifth neighbour. Results of the $14~000$ BEM simulations are shown in Fig. \ref{fig:BEMresults}. We observe that even at the smallest volume fraction there is still small but noticeable spread of the drag force magnitude, proving, that in case of Stokes flow, the Stokes drag model only truly holds for cases of isolated spheres. Given that the mean interparticle distance scales as $\varphi^{-1/3}$, the long-range disturbance of the stress field extends over a region on the order of $\mathcal{O}(10^2)$ particle diameters into the surrounding fluid. However, this picture changes in the case of small particles suspended in turbulent flow, where the extent of stress field disturbance is limited by the size of the turbulent eddies, and the fluid dynamics are instead dominated by turbulent stresses.

\begin{figure}
    \centering
    \begin{tikzpicture}
        \begin{axis}[
            xlabel={$\varphi$ (-)},
            ylabel={$\Re_\text{p} c_\text{D}$ (-)},
            zlabel={Probability density (-)},
            grid=major,
            width = 0.45\textwidth,
            xmode=log,
            ymin = 10,
            ymax = 24,
            zmin =  0,
            zmax = 1,
            ztick distance=0.5,
            ytick distance=5,
            view={50}{30}
        ]

        \addplot3[
            smooth,
            color=green!50!black, 
            fill=green!50!black, 
            fill opacity=0.6
        ] 
        table[x index=2, y index=0, z index=1, col sep=comma] {plots/PDF_1e-9.csv} \closedcycle;
        
        \addplot3[
            smooth,
            color=green!50!black, 
            fill=green!50!black, 
            fill opacity=0.6
        ] 
        table[x index=2, y index=0, z index=1, col sep=comma] {plots/PDF_1e-8.csv} \closedcycle;

        \addplot3[
            smooth,
            color=green!50!black, 
            fill=green!50!black, 
            fill opacity=0.6
        ] 
        table[x index=2, y index=0, z index=1, col sep=comma] {plots/PDF_1e-7.csv} \closedcycle;

        \addplot3[
            smooth,
            color=green!50!black, 
            fill=green!50!black, 
            fill opacity=0.6
        ] 
        table[x index=2, y index=0, z index=1, col sep=comma] {plots/PDF_1e-6.csv} \closedcycle;

        \addplot3[
            smooth,
            color=green!50!black, 
            fill=green!50!black, 
            fill opacity=0.6
        ] 
        table[x index=2, y index=0, z index=1, col sep=comma] {plots/PDF_1e-5.csv} \closedcycle;

        \addplot3[
            smooth,
            color=green!50!black, 
            fill=green!50!black, 
            fill opacity=0.6
        ] 
        table[x index=2, y index=0, z index=1, col sep=comma] {plots/PDF_1e-4.csv} \closedcycle;

        \addplot3[
            smooth,
            color=green!50!black, 
            fill=green!50!black, 
            fill opacity=0.6
        ] 
        table[x index=2, y index=0, z index=1, col sep=comma] {plots/PDF_1e-3.csv} \closedcycle;

        \end{axis}
\end{tikzpicture}
\begin{tikzpicture}
    \begin{axis}[
            xlabel={$\varphi$ (-)},
            ylabel={$\Re_\text{p} c_\text{D}$ (-)},
            grid=major,
            width = 0.45\textwidth,
            xmode=log,
            ymin = -1.2,
            ymax = 1.2,
            zmin =  0,
            zmax = 8,
            ztick distance=2,
            ytick distance=0.5,
            view={50}{30}
        ]

        \addplot3[
            smooth,
            color=red, 
            fill=red, 
            fill opacity=0.6
        ] 
        table[x index=2, y index=0, z index=1, col sep=comma] {plots/fx_PDF_1e-9.csv} \closedcycle;
        
        \addplot3[
            smooth,
            color=red, 
            fill=red, 
            fill opacity=0.6
        ] 
        table[x index=2, y index=0, z index=1, col sep=comma] {plots/fx_PDF_1e-8.csv} \closedcycle;

        \addplot3[
            smooth,
            color=red, 
            fill=red, 
            fill opacity=0.6
        ] 
        table[x index=2, y index=0, z index=1, col sep=comma] {plots/fx_PDF_1e-7.csv} \closedcycle;

        \addplot3[
            smooth,
            color=red, 
            fill=red, 
            fill opacity=0.6
        ] 
        table[x index=2, y index=0, z index=1, col sep=comma] {plots/fx_PDF_1e-6.csv} \closedcycle;

        \addplot3[
            smooth,
            color=red, 
            fill=red, 
            fill opacity=0.6
        ] 
        table[x index=2, y index=0, z index=1, col sep=comma] {plots/fx_PDF_1e-5.csv} \closedcycle;

        \addplot3[
            smooth,
            color=red, 
            fill=red, 
            fill opacity=0.6
        ] 
        table[x index=2, y index=0, z index=1, col sep=comma] {plots/fx_PDF_1e-4.csv} \closedcycle;

        \addplot3[
            smooth,
            color=red, 
            fill=red, 
            fill opacity=0.6
        ] 
        table[x index=2, y index=0, z index=1, col sep=comma] {plots/fx_PDF_1e-3.csv} \closedcycle;


        \addplot3[
            smooth,
            color=blue,
            fill=blue,
            fill opacity=0.6
        ] 
        table[x index=2, y index=0, z index=1, col sep=comma] {plots/fz_PDF_1e-9.csv} \closedcycle;
        
        \addplot3[
            smooth,
            color=blue,
            fill=blue,
            fill opacity=0.6
        ] 
        table[x index=2, y index=0, z index=1, col sep=comma] {plots/fz_PDF_1e-8.csv} \closedcycle;

        \addplot3[
            smooth,
            color=blue, 
            fill=blue,
            fill opacity=0.6
        ] 
        table[x index=2, y index=0, z index=1, col sep=comma] {plots/fz_PDF_1e-7.csv} \closedcycle;

        \addplot3[
            smooth,
            color=blue, 
            fill=blue, 
            fill opacity=0.6
        ] 
        table[x index=2, y index=0, z index=1, col sep=comma] {plots/fz_PDF_1e-6.csv} \closedcycle;

        \addplot3[
            smooth,
            color=blue, 
            fill=blue,
            fill opacity=0.6
        ] 
        table[x index=2, y index=0, z index=1, col sep=comma] {plots/fz_PDF_1e-5.csv} \closedcycle;

        \addplot3[
            smooth,
            color=blue, 
            fill=blue,
            fill opacity=0.6
        ] 
        table[x index=2, y index=0, z index=1, col sep=comma] {plots/fz_PDF_1e-4.csv} \closedcycle;

        \addplot3[
            smooth,
            color=blue, 
            fill=blue,
            fill opacity=0.6
        ] 
        table[x index=2, y index=0, z index=1, col sep=comma] {plots/fz_PDF_1e-3.csv} \closedcycle;

        \end{axis}
    \end{tikzpicture}
    \caption{Probability density function (p.d.f.) for measured streamwise (left) and spanwise and vertical (right) non-dimensional force components at different particle volume fractions. The spanwise and vertical component plots are almost identical and hence plotted on top of each other.}
    \label{fig:PDF}
\end{figure}
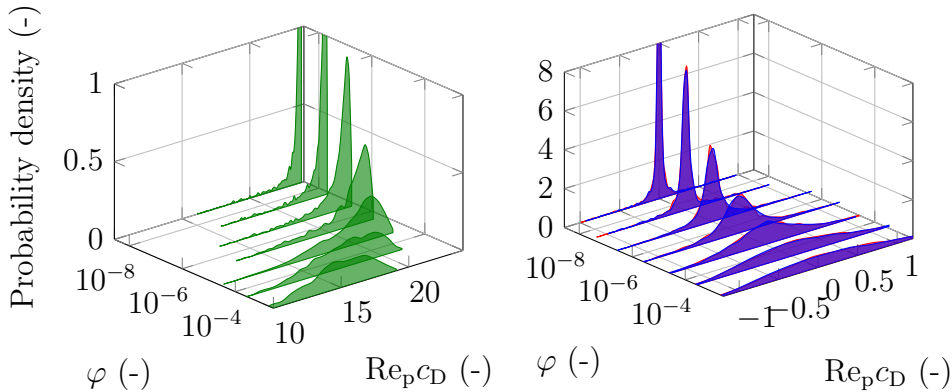

We can plot the data from Fig. \ref{fig:BEMresults} as probability density, presented in Fig. \ref{fig:PDF}. As we increase the particle volume fraction, the hydrodynamic interactions have a greater effect on reducing the streamwise component of the drag, and the lateral components become more pronounced. For the streamwise component of the force, the maximum probability is near the theoretical Stokes value, $\Re_\text{p} c_\text{D} = 24$. For $30\%$ of the measurements, the actual streamwise force is $15\%$ lower than the theoretical one. For $10\%$ of measured streamwise forces, this reduction is $34\%$. The vertical and the spanwise components exhibit approximately the same features. They loosely follow a Laplace distribution, with a mean value of $0$ and a variance of $0.126$ in case of the spanwise force and $0.124$ in case of the vertical component. Although studying a different particle regime ($\Re = 10, 0.1 < \varphi < 0.3 $), the study by \citet{jbaraPhysicsinspiredNeuralNetwork2025} showed a similar probability distribution of particle forces. This similarity likely stems from the fact that, even in the dilute Stokes regime, the long--range hydrodynamic interactions induced by neighbouring particles impose persistent fluctuations in the local stress field, producing force distributions that closely mirror those observed at higher $\Re$ numbers or volume fractions. To design an appropriate ANN, the statistical analysis of the dataset behaviour has a big importance when it comes to choosing the adequate loss function and activation function. The latter discussion is left for a later section.

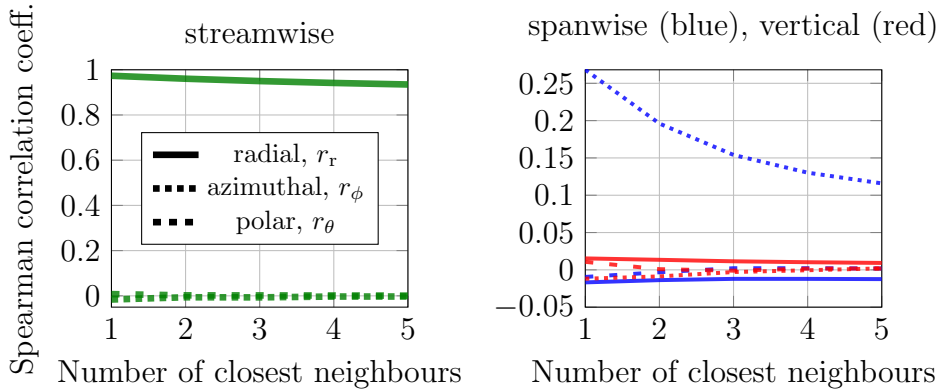
\begin{figure}[h]
    \centering
    \begin{tikzpicture}
        \begin{axis}
            [
                xlabel={Number of closest neighbours},
                ylabel={Spearman correlation coeff.},
                title={streamwise},
                enlargelimits=false,
                width = 0.4\textwidth,
                xmin = 1,
                xmax=5,
                grid=major,
                ymin=-0.05,
                ymax=1,
                ytick distance = 0.2,
                yticklabel style={
                /pgf/number format/fixed,
                /pgf/number format/precision=5
                },
                scaled y ticks=false,
                legend style={at={(0.1,0.5)},anchor=west},
                legend columns=1
            ]
            \addplot[color=black, solid, line width = 2.5pt,]coordinates {(-11,-11) (-1,-1)}; 
            \addlegendentry{\footnotesize radial, $r_\text{r}$}
            \addplot[color=black, dotted, line width = 2.5pt,]coordinates {(-11,-11) (-1,-1)}; 
            \addlegendentry{\footnotesize azimuthal, $r_\phi$}
            \addplot[color=black, dashed, line width = 2.5pt,]coordinates {(-11,-11) (-1,-1)}; 
            \addlegendentry{\footnotesize polar, $r_\theta$}

            \addplot[
                color=green!50!black, 
                opacity=0.8,
                line width = 2.5pt,
                solid
                ] 
            table[x index=0, y index=1, col sep=comma] {plots/silaY_correlation.csv};

            \addplot[
                color=green!50!black, 
                opacity=0.8,
                line width = 2.5pt,
                dotted
                ] 
            table[x index=0, y index=2, col sep=comma] {plots/silaY_correlation.csv};

            \addplot[
                color=green!50!black, 
                opacity=0.8,
                line width = 2.5pt,
                loosely dashed
                ] 
            table[x index=0, y index=3, col sep=comma] {plots/silaY_correlation.csv};

        \end{axis}
    \end{tikzpicture}
    \begin{tikzpicture}
        \begin{axis}
            [
                xlabel={Number of closest neighbours},
                legend style={draw=none, fill=none},
                title={spanwise (blue), vertical (red)},
                enlargelimits=false,
                width = 0.4\textwidth,
                grid=major,
                ymin=-0.05,
                ytick distance = 0.05,
                yticklabel style={
                /pgf/number format/fixed,
                /pgf/number format/precision=5
                },
                scaled y ticks=false,
                legend style={at={(0,0)},anchor=west},
                legend columns=1
            ]


            \addplot[
                color=blue, 
                opacity=0.8,
                line width = 1.5pt,
                solid
                ] 
            table[x index=0, y index=1, col sep=comma] {plots/silaX_correlation.csv};

            \addplot[
                color=blue, 
                opacity=0.8,
                line width = 1.5pt,
                dotted
                ] 
            table[x index=0, y index=2, col sep=comma] {plots/silaX_correlation.csv};

            \addplot[
                color=blue, 
                opacity=0.8,
                line width = 1.5pt,
                loosely dashed
                ] 
            table[x index=0, y index=3, col sep=comma] {plots/silaX_correlation.csv};

            \addplot[
                color=red, 
                opacity=0.8,
                line width = 1.5pt,
                solid
                ] 
            table[x index=0, y index=1, col sep=comma] {plots/silaZ_correlation.csv};

            \addplot[
                color=red, 
                opacity=0.8,
                line width = 1.5pt,
                dotted
                ] 
            table[x index=0, y index=2, col sep=comma] {plots/silaZ_correlation.csv};

            \addplot[
                color=red, 
                opacity=0.8,
                line width = 1.5pt,
                loosely dashed
                ] 
            table[x index=0, y index=3, col sep=comma] {plots/silaZ_correlation.csv};

        \end{axis}
    \end{tikzpicture}
    \caption{The plot of Spearman correlation coefficient with respect to the number of closest neighbours added into correlation analysis.}
    \label{fig:correlation}
\end{figure}

Fig. \ref{fig:correlation} shows the correlation of individual neighbour positions $\mathbf{r}$ in spherical coordinates with $\underline{r} = [r_\text{r}, r_\phi, r_\theta]$. The horizontal axis represents the number of closest neighbours included in the correlation analysis. E.g., number $1$ on the horizontal axis denotes that only one closest neighbour was included into correlation estimate, while number $2$ on the horizontal axis means two closest neighbours were included, and their correlation value was averaged. For correlation analysis, the \citet{spearmanProofMeasurementAssociation1904} correlation factor was chosen for assessing non--linear dependence of variables. Intuitively, the radial component of neighbour coordinates will have the most effect on the streamwise reference particle force component, coinciding with similar observations by \citet{akikiForceVariationArrays2016}. Again, we attribute this to the far-reaching nature of the diffusion--driven momentum transfer. Interestingly, a moderate correlation is observed between the azimuthal angle and the spanwise force component. This correlation diminishes as we increase the number of nearest neighbours in the calculation. Since we observe no significant correlation between the polar angle and any of the force components, we can roughly justify the axisymmetric assumption of the problem, posed by some other authors \citep{akikiPairwiseinteractionExtendedPointparticle2017, mooreHybridPointparticleForce2019, balachandarParticleresolvedAccuracyEuler2020}. Interestingly, poor correlation ($\mathcal{O}(10^{-3})$) for all variables is observed, if we input the neighbour positions in Cartesian coordinates. The weak overall correlation between the coordinates and the forces highlights the complexity of the problem, emphasizing its multibody nature, which cannot be adequately described by simple non-linear relationships, meaning a multi-variable, non-linear model must be established to deterministically compute the reference particle forces.

\section{Deterministic model}\label{sect:deterministicModel}
The dataset, obtained from the $14~000$ BEM simulations and presented in the previous section, was used to train an ANN, as this regression approach proved to be able to solve similar problems \citep{heSupervisedMachineLearning2019,balachandarParticleresolvedAccuracyEuler2020,seyed-ahmadiPhysicsinspiredArchitectureNeural2022,siddaniPointparticleDragLift2023,jbaraPhysicsinspiredNeuralNetwork2025}. Here, it is in place to introduce the term "order of interaction", in our context meaning the number of neighbouring particles involved in a given interaction with the reference particle at a time. All standard streamwise drag models, given as $F_\text{d} = F_\text{d} (\Re_{\text{p}}, \varphi)$, represent the $0^{\text{th}}$ order interaction. The essence of the PIEP approach \citep{akikiPairwiseInteractionExtended2017} is the expansion of the drag model with a $1^{\text{st}}$ order interaction term, that accounts for the interactions of one neighbour at a time, assuming that their effects are independent and can be superposed. Analogously, increasing the order of interactions incorporates the influence of more neighbours simultaneously, progressively capturing the multibody nature of the problem with greater accuracy. 
We propose a novel, physics-based architecture, built specifically for solving such multibody type problems, based on the following findings:
\begin{itemize}
    \item similarly to \citet{seyed-ahmadiPhysicsinspiredArchitectureNeural2022}, we integrate the PIEP into our ANN architecture, as this approach is physically justified and also solves the "curse of dimensionality", that comes by accounting for all variables at the same time. Furthermore, inputting the coordinates into the ANN as a flat vector eliminates the crucial information regarding the association of each coordinate with a specific particle. The PIEP approach fixes this issue at the expense of not accounting for the complete multibody effects.
    \item Since it was found by other authors \citep{batchelorSedimentationDiluteDispersion1972,seyed-ahmadiMicrostructureinformedProbabilitydrivenPointparticle2020}, and confirmed by our own observations (Fig. \ref{fig:correlation}), that higher--order interactions play a dominant role with closest neighbours, we account for a $2^{\text{nd}}$ order interaction with two closest neighbours, therefore capturing the multibody effect at the closest level.
    \item Correlation analysis demonstrated that spherical coordinates correlate with specific force components; therefore, we used them as inputs instead of the Cartesian coordinates.
    \item Following the same strategy as the previously mentioned authors, we train an ANN for each force component separately.
\end{itemize}

\subsection{Novel Interaction-Decomposed Neural Network (IDNN) architecture and algorithm}
In the first part of the following section, the working paradigm of an ANN will be briefly outlined. The proposed novel IDNN architecture for solving hydrodynamic multibody interactions is presented afterwards. Although the foundational principles of ANNs have been established for decades, their true capabilities have only been fully realized in the past ten years, largely owing to advances in GPU-accelerated computing. While the physical system's emergent behaviour will be examined in detail in subsequent sections, we have deliberately refrained from venturing into the intricate mathematical formalism underlying ANNs. An attempt to outline the multi-parametric nonlinear regression framework would risk oversimplification and detract from the scientific rigor. Readers seeking comprehensive theoretical insights are instead referred to sources, such as \citet{adampaszkeAutomaticDifferentiationPyTorch2017}.

An ANN performs a series of linear and non--linear operations on the given inputs. One of its main advantages is the flexibility in terms of dimensionality, meaning that an arbitrary $N$ dimensional set of features could be mapped onto an $M$ dimensional space, $f: \mathbb{R}^N \rightarrow \mathbb{R}^M$. The aforementioned linear and non--linear transformations are sequentially captured in the mapping process as
\begin{equation}\label{eq:forwardPass}
    \underline{h}^{[l+1]} = \sigma \left( \underline{W}^{[l+1]} \underline{h}^{[l]} + \underline{b}^{[l+1]} \right),
\end{equation}
where $\underline{h}^{[l]}$ denotes the matrix of the $[l]$--th hidden layer and $\underline{W}$ and $\underline{b}$ denote the weights and biases respectively. Needless to say, that in our case, for the input layer, the matrix coefficients are the coefficients of the position vector, $\underline{h}^{[0]} \equiv \underline{r}$, and analogously for the output layer, $\underline{h}^{[l_{max}]} \equiv \underline{F}$. The non-linear operation is denoted as $\sigma$, and represents the so-called activation function.

\begin{figure}[t]
    \centering
        \begin{tikzpicture}[>=stealth, thick, scale=0.3]
        \footnotesize
        \coordinate (A) at (-2,0);
        \coordinate (B) at (2.1,0.5);
        \coordinate (C) at (1.1,2.2);
        \coordinate (D) at (4,1.9);
        \coordinate (E) at (1,0.8);
        \coordinate (F) at (-1.2,-0.4);
        
        \draw[ball color=green!60] (A) circle (10pt) node[above, yshift=5pt] {$\mathbf{r}_5$};
        \draw[ball color=blue!60] (B) circle (10pt) node[above, yshift=5pt] {$\mathbf{r}_1$};
        \draw[ball color=blue!60] (C) circle (10pt) node[above, yshift=5pt] {$\mathbf{r}_2$};
        \draw[ball color=green!60] (D) circle (10pt) node[above, yshift=5pt] {$\mathbf{r}_4$};
        \draw[ball color=red!70!blue] (E) circle (10pt);
        \draw[ball color=green!60] (F) circle (10pt) node[above, yshift=5pt] {$\mathbf{r}_3$};
    \end{tikzpicture} \\
    
    \begin{tikzpicture}[
        neuronInput/.style={circle, draw, minimum size=0.1cm},
        neuronInput2/.style={circle, dashed, draw, minimum size=0.1cm},
        squareInput/.style={rectangle, draw, minimum width=1cm, minimum height=0.1cm},
        neuronHidden/.style={circle, draw, minimum size=0.05cm},
        input/.style={squareInput},
        hidden/.style={neuronHidden, fill=gray!30},
        output/.style={neuronInput, fill=red!30},
        output2/.style={neuronInput, fill=red!30},
        output3/.style={neuronInput, fill=red!30},
        connections/.style={->, gray!80, -stealth},
        connections2/.style={->, gray!80, -stealth},
        scale = 0.8
    ]
    \scriptsize
    \node at (3.3, 0) {$\mathbf{2^{\text{nd}}}$ \textbf{order block}};
    \node at (9.8, 0) {$\mathbf{1^{\text{st}}}$ \textbf{order block}};

    \draw [fill=blue!20, opacity=0.8, rounded corners] (1.8,-4.8) rectangle ++ (4.4,4.5);
    \draw [fill=green!20, opacity=0.8, rounded corners] (8.4,-4.8) rectangle ++ (4.4,4.5);

    \node[input, rotate=90] (I1) at (0.8,-1.2) {$\underline{r}_1 (r_1, \phi_1, \theta_1)$};
    \node[input, rotate=90] (I2) at (0.8,-4) {$\underline{r}_2 (r_2, \phi_2, \theta_2)$};

    \foreach \i in {1,...,2}
        \node[hidden] (H1\i) at (2.5,-\i*1.2) {$h^{[1]}_{\i}$};
    \node at (2.5,-3*1.2 + 0.5) {\vdots};
    \node[hidden] (H13) at (2.5,-4) {$h^{[1]}_{6}$};

    \foreach \i in {1,...,2}
        \node[hidden] (H2\i) at (4.2,-\i*1.2) {$h^{[2]}_{\i}$};
    \node at (4.2,-3*1.2 + 0.5) {\vdots};        
    \node[hidden] (H23) at (4.2,-4) {$h^{[2]}_{n}$};

    \foreach \i in {1,...,2}
        \node[hidden] (H3\i) at (5.5,-\i*1.2) {$h^{[l]}_{\i}$};
    \node at (5.5,-3*1.2 + 0.5) {\vdots};
    \node[hidden] (H33) at (5.5,-4) {$h^{[l]}_{n}$};

    % ...

    \node at (4.85,-4) {\dots};        
    \node at (4.85,-2.4) {\dots};        
    \node at (4.85,-1.2) {\dots};        

    % ...


    \node[output] (O) at (6.9,-1.2) {$\underline{F}_{12}$};
    \node[output2] (O2) at (7.4,-3.4) {$\underline{F}_3$};
        \node at (7.4,-4.05) {\vdots};
    \node[output3] (O3) at (7.4,-5) {$\underline{F}_5$};

    \foreach \i in {1,...,2}
        \node[hidden] (H4\i) at (9.1,-\i*1.2) {$h^{[l]}_{\i}$};
    \node at (9.1,-3*1.2 + 0.5) {\vdots};        
    \node[hidden] (H43) at (9.1,-4) {$h^{[l]}_{n}$};

    \node at (9.75,-4) {\dots};      
    \node at (9.75,-2.4) {\dots};        
    \node at (9.75,-1.2) {\dots};

    \foreach \i in {1,...,2}
        \node[hidden] (H6\i) at (10.4,-\i*1.2) {$h^{[2]}_{\i}$};
    \node at (10.4,-3*1.2 + 0.5) {\vdots};        
    \node[hidden] (H63) at (10.4,-4) {$h^{[2]}_{n}$};

    \foreach \i in {1,...,2}
        \node[hidden] (H7\i) at (12.1,-\i*1.2) {$h^{[1]}_{\i}$};
    \node[hidden] (H73) at (12.1,-4) {$h^{[1]}_{3}$};

    \node[input, rotate=90] (I3) at (13.9,-1) {$\underline{r}_3 (r_3, \phi_3, \theta_3)$};
    \node at (13.9,-2.35) {\vdots};
    \node[input, rotate=90] (I4) at (13.9,-4) {$\underline{r}_5 (r_5, \phi_5, \theta_5)$};

    \foreach \i in {1,2}
        \foreach \j in {1,...,3}
            \draw[connections] (I\i.south) -- (H1\j.west);

    \foreach \i in {1,2}
        \foreach \j in {1,...,3}
            \draw[connections] (I3.north) -- (H7\j.east);

    \foreach \i in {1,...,3}
        \foreach \j in {1,...,3}
            \draw[connections] (H1\i.east) -- (H2\j.west);

    \foreach \i in {1,...,3}
        \draw[connections] (H3\i.east) -- (O);



    \foreach \i in {1,...,3}
        \draw[connections2] (H4\i.west) -- (O2);

    \foreach \i in {1,...,3}
        \foreach \j in {1,...,3}
            \draw[connections] (H7\i.west) -- (H6\j.east);

    \node at (7.4,-6) {$\underline{F}_{\text{IDNN}} = \underline{F}_{12} + \underline{F}_{3} + \underline{F}_{4} + \underline{F}_{5}$};
\end{tikzpicture}
    \caption{Schematic representation of a single Interaction-Decomposed Neural Network (IDNN). Separate IDNNs are employed to predict the streamwise force component and the combined lateral components (spanwise and vertical). As a result, the output neuron is denoted in bold to indicate that it can represent either a scalar value, $\underline{F}_{IDNN} \in \mathbb{R}$, for the streamwise component, or a two-dimensional vector, $\underline{F}_{IDNN} \in \mathbb{R}^2$, for the lateral components. The IDNN features: (i) a two-block architecture, with a $1^{\text{st}}$ order block for pairwise interactions and a $2^{\text{nd}}$ order block for higher-order effects; and (ii) input sorting by radial distance, implicitly encoding the relative influence of each neighbour (see Fig. \ref{fig:correlation}). For the above case, the following holds: $|\mathbf{r}_1| < |\mathbf{r}_2| < |\mathbf{r}_3| < |\mathbf{r}_4| < |\mathbf{r}_5|$, while the reference particle is shown in red}\label{fig:nnModel}
\end{figure}
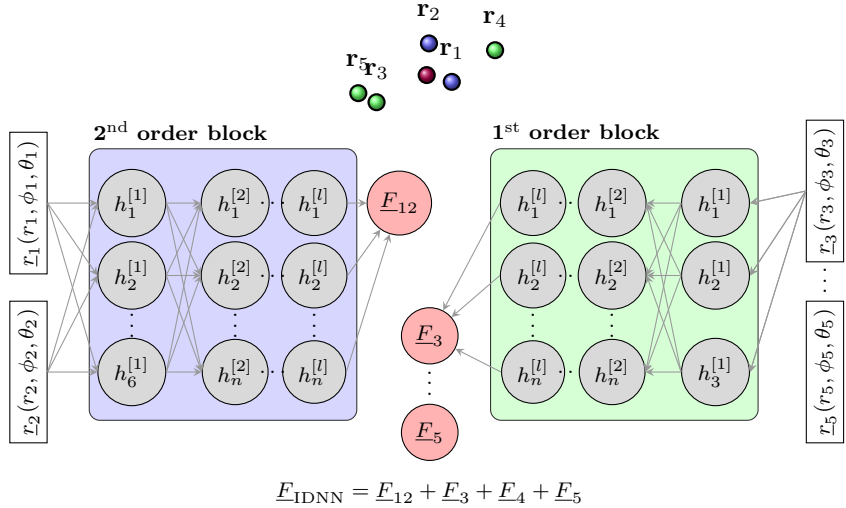

The proposed IDNN architecture is shown in Fig. \ref{fig:nnModel}. The IDNN is divided into two blocks; the $2^{\text{nd}}$ order block, that captures $2^{\text{nd}}$ order interactions, and $1^{\text{st}}$ order block, that captures the $1^{\text{st}}$ order, or pairwise interactions. Prior to training, we apply a sorting process to the whole dataset, so that neighbour $\underline{r}_1$ will always be the closest neighbour in a given particle arrangement, and neighbour $\underline{r}_5$ will be the farthest. Separate IDNNs are employed to predict the streamwise force component and the combined lateral components (spanwise and vertical). As a result, the output neuron is underlined to indicate that it can represent either a scalar value, $\underline{F}_{IDNN} \in \mathbb{R}$, for the streamwise component, or a two-component matrix, $\underline{F}_{IDNN} \in \mathbb{R}^2$, for the lateral components. The proposed training algorithm goes as follows:
\begin{enumerate}
    \item first, the spherical coordinates of the two closest neighbours are simultaneously passed into the $2^{\text{nd}}$ order block to obtain the prediction of the $2^{\text{nd}}$ order contribution.
    \item Next, the remaining three neighbours are passed one at a time into the $1^{\text{st}}$ order block to obtain the prediction of the $1^{\text{st}}$ order contribution. These neighbours are also passed in order of increasing distance.
    \item The contribution from the $2^{\text{nd}}$ order block and the three contributions from the $1^{\text{st}}$ order block are superposed, to obtain the force prediction in the current epoch.
    \item The loss function is computed.
    \item Backpropagation is performed, first for the $2^{\text{nd}}$ order block and second for the $1^{\text{st}}$ order block, to adjust the corresponding IDNN parameters.
    \item The training loop is repeated.
\end{enumerate}

The choice of some IDNN hyperparameters, such as the activation function and the loss function, must be influenced by the physical system we are trying to predict. Other, hyperparameters, such as number of hidden layers and neurons, as well as the learning rate, etc., are a matter of hyperparameter tuning \citep{neumannHeuristicHyperparameterOptimization2018}. For regression tasks, the mean-squared error (MSE), is said to be the best practice \citep{seyed-ahmadiPhysicsinspiredArchitectureNeural2022}. We compute it as
\begin{equation}
    L_{\text{MSE}} = \frac{1}{N_{\text{train}}} \sum_{i}^{N_{\text{train}}} \left[ \underline{F}_{i, \text{IDNN}} - \underline{F}_{i, \text{BEM}} \right]^2,
\end{equation}
where index $i$ denotes the training sample from the dataset. Here $N_{\text{train}}$ is the size of the training dataset. In our case, $N_{\text{train}}$ accounts to $90\%$ of the whole dataset, as $10\%$ was randomly taken to be the validation dataset and was not included in training. During training, a validation forward pass is performed every $100$ epochs, to test the generalizability of the model by testing it on unseen data. The performance (validation error) of the validation data was also evaluated by the MSE. The validation error also serves as a stopping criterion; when it begins to increase while the training loss continues to decrease, it indicates overfitting. This is the point at which training should be stopped to prevent the model from memorizing the training data instead of generalizing well. 
The choice of the activation function is motivated by the behaviour of the dataset. For the streamwise force component (Fig. \ref{fig:PDF}, left), the force is bounded by the theoretical Stokes force ($\Re_\text{p} c_\text{D} = 24$), and the ReLU activation function appears to effectively capture this behaviour. For the lateral force components (Fig. \ref{fig:PDF}, right), the data points fluctuate around a mean value of zero, a behaviour that appears to be well captured by the hyperbolic tangent (Tanh) activation function. A similar choice was also made by \citet{siddaniPointparticleDragLift2023}. For optimizing the IDNN parameters, the AdamW optimizer was used.
The output layer was set to have a linear activation function and no bias, similar to \citet{seyed-ahmadiPhysicsinspiredArchitectureNeural2022}.

\subsection{Integration of the IDNN into the CFD framework}
Once the IDNN model is trained, we can use it to predict forces on particles by inputing the neighbouring coordinates in a way described above. Most machine learning libraries use a black-box approach for deploying trained models, allowing users to save a model and load it in a different instance without needing to understand its internal workings. However, in our case, the model must be integrated directly into the Lagrangian particle tracking solver, meaning we cannot rely on standard black-box deployment. Instead, we need a more transparent implementation that allows interaction with the solver's data structures and solver framework. In our case, we integrate the trained model into the open-source CFD package OpenFOAM v11 \citep{wellerTensorialApproachComputational1998}, that makes use of the object-oriented programming paradigm of the C\texttt{++} language. The trained model's parameters--weight matrices and bias vectors--are read and saved by the OpenFOAM Lagrangian solver prior to the Larangian time loop. For each particle in the system, the five nearest neighbours are found in each time-step, and their coordinates sequentially passed through the IDNN as in Eq. (\ref{eq:forwardPass}), integrated into the Lagrangian solver. The obtained force is then explicitly added into the Maxey-Riley equation, that dictates the movement of the particle. To obtain the particle force, the trained IDNN model can be regarded as an alternative force model. Instead of relying on a single mathematical expression, the computational complexity of this approach is determined by the IDNN's architecture, specifically the number of parameters and operations required for inference, which in our case amounts to $\mathcal{O} (10^3)$ operations, and does not pose significant burden to computational power. The state-of-the-art in numerical modelling is increasingly shifting towards the use of ANN-trained models as surrogates for traditional models. Thereby, with more complex models, more mathematical operations will be required for inference. In the present case, the majority of the computational cost is used on the five nearest neighbours search, which must also allow for parallelization, to make the IDNN model useful for CFD. For that matter, the OpenFOAM's so--called "interaction list" framework was adopted, which allows for particle access across processor boundaries. OpenFOAM's interaction lists are originally implemented in the framework of particle--particle collisions \citep{wedelNovelParticleParticle2024}, that also require particle data sharing across processor boundaries.
\section{Particle-laden flow governing equations}
We model the particle-laden flow using point-particles, meaning that we do not spatially resolve the particle. The particle kinematics for the $i$-th particle are solved using the Lagrangian method,
\begin{equation}\label{eq:forceIDNN}
    m_{\text{p}, i} \frac{\text{d} \mathbf{v}_{\text{p}, i}}{\text{d} t} = \mathbf{F}_{\text{g}, i} + \mathbf{F}_{\text{b}, i} + \mathbf{F}_{\text{IDNN}, i},
\end{equation}
where $\mathbf{F}_{\text{g}}$ and $\mathbf{F}_{\text{b}}$ are the gravity and buoyant forces. The force $\mathbf{F}_{\text{IDNN}}$ is predicted by the IDNN, integrated into the Lagrangian solver. The BEM database was constructed in the particle's local coordinate system (LCS), therefore appropriate coordinate transformations must be applied prior to performing inference with the integrated IDNN within the Lagrangian solver. The detailed transformations are given in appendix \ref{appB}. In short, we obtain the reference particle coordinates, as well as the neighbour coordinates, $\underline{r}_{i}$, $\underline{r}_{i,j}$ in the global coordinate system (GCS). The IDNN requires the neighbour particle coordinates in the LCS, $\underline{r}'_{j}$, in order to perform a correct forward pass as shown in Eq. (\ref{eq:GCS-LCS}). We perform a forward pass through the IDNN and obtain the prediction of the reference particle force,
\begin{equation}
    \mathbf{F}_{\text{IDNN}, i} = \mathcal{F} \left( \mathbf{r}_{i, j = 1} \dots \mathbf{r}_{i, j = M} \right), \hspace{0.5cm} M=5
\end{equation}
where $\mathcal{F}$ is the abstract notation of Eq. (\ref{eq:forwardPass}), that describes the linear and non-linear transformations performed during the forward pass, applied to $M$ neighbour particle coordinates. The obtained particle force is then used in Eq. (\ref{eq:forceIDNN}) in an explicit manner, to contribute to the movement of Lagrangian particles.
\section{IDNN training results}
To identify a near-optimal configuration of hidden layers and neurons for our approach, we conducted multiple training runs and monitored the MSE for each setup. The number of hidden layers and neurons in the IDNN was systematically varied, and $10$--fold cross-validation was applied to each configuration. While this exhaustive search strategy is not commonly used in other contexts due to computational demands, it was feasible in our case because the IDNN model is relatively simple, and each training run requires minimal computational time—unlike, for instance, large language models where such approach would be impractical \citep{almanFineGrainedComplexityGradient2024}.
The proposed IDNN is trained separately for the streamwise and the lateral components. This means that the IDNN for the streamwise component performs mapping $\mathbb{R}^6 \rightarrow \mathbb{R}$ for the $2^{\text{nd}}$ order block and three times $\mathbb{R}^3 \rightarrow \mathbb{R}$ for the $1^{\text{st}}$ order block, while the IDNN for the lateral components performs mapping $\mathbb{R}^6 \rightarrow \mathbb{R}^2$, three times $\mathbb{R}^3 \rightarrow \mathbb{R}^2$ for the $2$nd and $1^{\text{st}}$ order blocks respectively. The optimal hidden layer configuration, yielding the lowest mean MSE, was found to be $40$-$40$ for the $2^{\text{nd}}$ order block and $30$-$30$-$30$ for the $1^{\text{st}}$ order block. This configuration was used consistently for training both the streamwise and lateral components. The whole training process was conducted using the PyTorch open-source machine learning library \citep{adampaszkeAutomaticDifferentiationPyTorch2017} on Nvidia RTX A5000 GPU hardware.

Determining a stopping criterium for the training of an ANN is not as straightforward as it would be for numerical computing, for example when numerically solving systems of partial differential equations (PDEs). The goal of numerical methods is to iteratively find a solution, until the error between iterations falls below a certain threshold, meaning that an approximation of the true solution has been found. We can also detect this by confirming that the numerical solution has reached a plateau. In ANN training, reaching a plateau in the validation error does not necessarily indicate global convergence, as lower minima may exist. Therefore, stopping criteria must be carefully tailored to the optimization problem. In our approach, we adopt the following early stopping method:
\begin{enumerate}
    \item We monitor the validation error every $100$ training epochs.
    \item If the validation error decreases compared to the previously recorded minimum, we save the new best model; otherwise, training continues.
    \item If the validation error does not improve for a predefined number of consecutive training epochs, training is stopped to prevent overfitting.
    \item The predefined patience threshold is chosen based on empirical analysis to balance convergence and generalization.
\end{enumerate}
This stopping method ensures that training halts only when further improvement is unlikely, while also avoiding premature termination. In our case, we set the patience threshold to $10~000$ training epochs.

To evaluate the performance of the IDNN, we employ the coefficient of determination ($R^2$), which quantifies the degree of linear correlation between the predicted forces and the ground-truth values. We tested the IDNN against the classic dense ANN, as used by \citet{heSupervisedMachineLearning2019, balachandarParticleresolvedAccuracyEuler2020}.
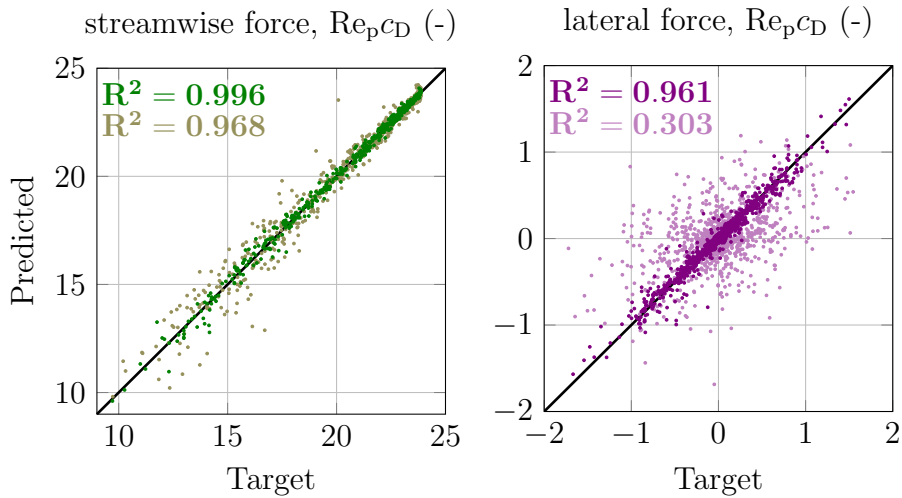
\begin{figure}[h]
    \centering
    \begin{tikzpicture}
        \begin{axis}
            [
                title={streamwise force, $\Re_\text{p} c_\text{D}$ (-)},
                xlabel={ Target},
                ylabel={Predicted},
                grid=major,
                width = 0.45\textwidth,
                height = 0.45\textwidth,
                xmin =9,
                xmax=25,
                ymax = 25,
                ymin=9,
                each nth point=10
            ]

            \addplot
            [
                mark=none,
                line width=1pt,
                domain=9:30,
                color=black
            ]
            {x};

            \addplot[
                only marks, 
                mark=*, 
                color=yellow!50!black, 
                mark size=0.5pt
            ] 
            table [col sep=comma] {plots/force_y_validation_dense_unsorted.dat};
            \node[yellow!50!black] at (rel axis cs:0.25,0.84) {$\mathbf{R^2 = 0.968}$};


            \addplot[
                only marks, 
                mark=*, 
                color=green!50!black, 
                mark size=0.5pt
            ] 
            table [col sep=comma] {plots/force_y_validation.dat};
            \node[green!50!black] at (rel axis cs:0.25,0.93) {$\mathbf{R^2 = 0.996}$};

        \end{axis}
    \end{tikzpicture}
    \begin{tikzpicture}
        \begin{axis}
            [
                title={lateral force, $\Re_\text{p} c_\text{D}$ (-)},
                xlabel={ Target},
                grid=major,
                width = 0.45\textwidth,
                height = 0.45\textwidth,
                xmin =-2,
                xmax=2,
                ymax = 2,
                ymin=-2,
                each nth point=10
            ]

            \addplot
            [
                mark=none,
                line width=1pt,
                domain=-5:5,
                color=black
            ]
            {x};

            \addplot[
                only marks, 
                mark=*, 
                color=blue!50!red!50!white, 
                mark size=0.5pt
            ] 
            table [col sep=comma] {plots/force_xz_validation_dense_unsorted.dat};
            \node[color=blue!50!red!50!white, ] at (rel axis cs:0.25,0.84) {$\mathbf{R^2 = 0.303}$};
            

            \addplot[
                only marks, 
                mark=*, 
                color=blue!50!red, 
                mark size=0.5pt
            ] 
            table [col sep=comma] {plots/force_xz_validation.dat};
            \node[blue!50!red] at (rel axis cs:0.25,0.93) {$\mathbf{R^2 = 0.961}$};

        \end{axis}
    \end{tikzpicture}
    \caption{
        $R^2$ metric for streamwise force (left panel): \tikzcircle[green!50!black, fill=green!50!black]{2pt} IDNN, \tikzcircle[yellow!50!black, fill=yellow!50!black]{2pt}  Dense ANN; and for lateral force (vertical + spanwise) (right panel): \tikzcircle[red!50!blue, fill=red!50!blue]{2pt} IDNN, \tikzcircle[red!50!blue!50!white, fill=red!50!blue!50!white]{2pt} Dense ANN
    }
    \label{fig:Rsquared}
\end{figure}

The results show that by implementing the IDNN we are able to obtain predictions that improve upon the dense ANNs ($R^2 = 0.92$ in case of \citet{heSupervisedMachineLearning2019}, $R^2 = -0.30$ to $-0.12$ in case of \citet{balachandarParticleresolvedAccuracyEuler2020}) as well as the pure PIEP-ANN results ($R^2 = 0.53 - 0.75$ in case of \citet{balachandarParticleresolvedAccuracyEuler2020}, $R^2 = 0.67 - 0.70$ in case of \citet{seyed-ahmadiPhysicsinspiredArchitectureNeural2022}, $R^2 = 0.40 - 0.73$ in case of \citet{mooreHybridPointparticleForce2019}). However, it is important to note that the physical regime considered in this study differs from those investigated by other authors, who typically focus on denser systems characterized by higher Reynolds numbers. As a result, direct comparisons may not be entirely fair, as the nature of the underlying physics significantly differs. In our case, the hydrodynamic interactions are primarily governed by diffusive momentum transfer, which may present a less complex challenge. Since Stokes flow is described by a linear PDE, the relatively poor performance of some predictive methods in other studies could be attributed to the presence of nonlinear inertial effects, which can hinder both the training process and the generalization capability of the models.
\section{Mineral dust sedimentation}
The trained IDNN model was used to simulate the sedimentation of particles in a quiescent fluid. We chose to simulate the sedimentation of fine mineral dust material, for which we use the monodisperse assumption, meaning that all particles have the same physical properties. To evaluate the statistical independence of the results from the number of particles, we conducted simulations in a cubic domain of varying side length $L$ with three different particle counts: $n=1000$, $n=10~000$ and $n=33~000$. We define the initial particle volume fraction as $\varphi_0 = n V_p / L^3$, where $V_p$ is the volume of a single particle. In all cases, particles were randomly distributed within the domain and allowed to settle under gravity. To maintain a constant number of particles across different values of $\varphi_0$, the domain size $L$ was adjusted accordingly. Throughout the sedimenting process, the local volume fraction surrounding each particle was evaluated using the distance to its five nearest neighbours, as defined in Eq. (\ref{eq:volFrac_definition}). There is no flow in the domain and the domain walls are periodic. The parameters of the simulation are given in Tab. \ref{tab:physicalProperties}.
\begin{table}
    \begin{center}
  \def~{\hphantom{0}}
  \begin{tabular}{lr}
    \textbf{Parameter} & \textbf{Value} \\
    Air Density (\(\rho_\text{f}\)) & $1.2041$ kg/m\(^3\) \\
    Dynamic Viscosity of air (\(\eta_\text{f}\)) & $1.81 \cdot 10^{-5}$ Pa.s \\
    Particle Diameter ($d_\text{p}$) & $25$~$\mu$m \\
    Particle Density (\(\rho_\text{p}\)) & $2500$ kg/m\(^3\) \\
    Characteristic particle response time for a particle  ($\tau$) & $4.79589$~ms \\
    Stokes settling velocity (\(v_\text{s, Stokes}\)) & $0.0470$~m/s \\
    \end{tabular}
    \caption{Physical properties used for the particle sedimentation simulation, corresponding to mineral dust, sedimenting in $20^{\circ} \text{C}$ air.}
    \label{tab:physicalProperties}
    \end{center}
\end{table}  

The theory behind settling velocity involves the balance of forces acting on a particle in a fluid. Assuming a very low particle Reynolds number, the settling velocity ($v_\text{s}$) for a spherical particle follows from Stokes drag law. The forces acting on the particle are the gravitational force
\begin{equation}
    \mathbf{F}_\text{g} = \frac{1}{6} \pi d_\text{p}^3 \rho_\text{p} \mathbf{g},
\end{equation}
the buoyancy force,
\begin{equation}
    \mathbf{F}_\text{b} = \frac{1}{6} \pi d_\text{p}^3 \rho_\text{f} \mathbf{g},
\end{equation}
and the force $\mathbf{F}_{\text{IDNN}}$ obtained through the IDNN inference. Since the focus is on sedimentation, the settling velocity will be determined using the streamwise component of the predicted force vector, $\mathbf{F}_{\text{IDNN}}$, which aligns with the direction of gravitational acceleration. From this point forward, the drag force will be denoted as $F_{\text{d}}$, and will specifically refer to the drag associated with the settling velocity. The classical Stokes drag reads as \citep{stokesEffectInternalFriction1851},
\begin{equation}\label{eq:StokesDrag}
    F_\text{d, Stokes} = 3 \pi \eta_f d_\text{p} v_\text{s, Stokes}.
\end{equation}
The drag force, obtained through $\mathbf
{F}_{\text{IDNN}}$ can be written as
\begin{equation}\label{eq:dragWithBeta}
    F_\text{d} = \mathbf{F}_{\text{IDNN}} \cdot \frac{\mathbf{g}}{|\mathbf{g}|} = \beta F_\text{d, Stokes},
\end{equation}
where an additional factor $\beta$ was introduced, which describes the change in Stokes drag due to the presence of other particles and $v_\text{s, Stokes}$ is the theoretical Stokes settling velocity for a particle with given physical parameters. The factor $\beta$ is equal to one for very low volume fractions and decreases with increasing volume fraction. The factor $1/\beta$ can be thought of as the non-dimensional settling velocity, represented by the ratio between the Stokes drag to the actual drag on a particle at a given volume fraction. We can make the force balance for the theoretical Stoke settling velocity as well as our modified settling velocity:
\begin{equation}\label{eq:settlingVelocity}
    v_\text{s, \text{Stokes}} = \frac{d_\text{p}^2 |\mathbf{g}| [\rho_\text{p} - \rho_\text{f}]}{18 \eta}, \hspace{0.5cm} v_\text{s} = \frac{d_\text{p}^2 |\mathbf{g}| [\rho_\text{p} - \rho_\text{f}]}{\beta 18 \eta}
\end{equation}
from which we can obtain the $1/\beta$ factor as the settling velocity at a given volume fraction, relative to the Stokes settling velocity,
\begin{equation}
    \frac{v_\text{s} (\varphi)}{v_{\text{s, Stokes}}} = \frac{1}{\beta}.
\end{equation}
Note that the measured settling velocity is written here as a function of the volume fraction, where in our case, the measured particle settling velocity is actually a direct function of the positions of the five neighbouring particles, which then inherently makes the measured settling velocity a function of the volume fraction.

\begin{figure}
    \centering
    \includegraphics{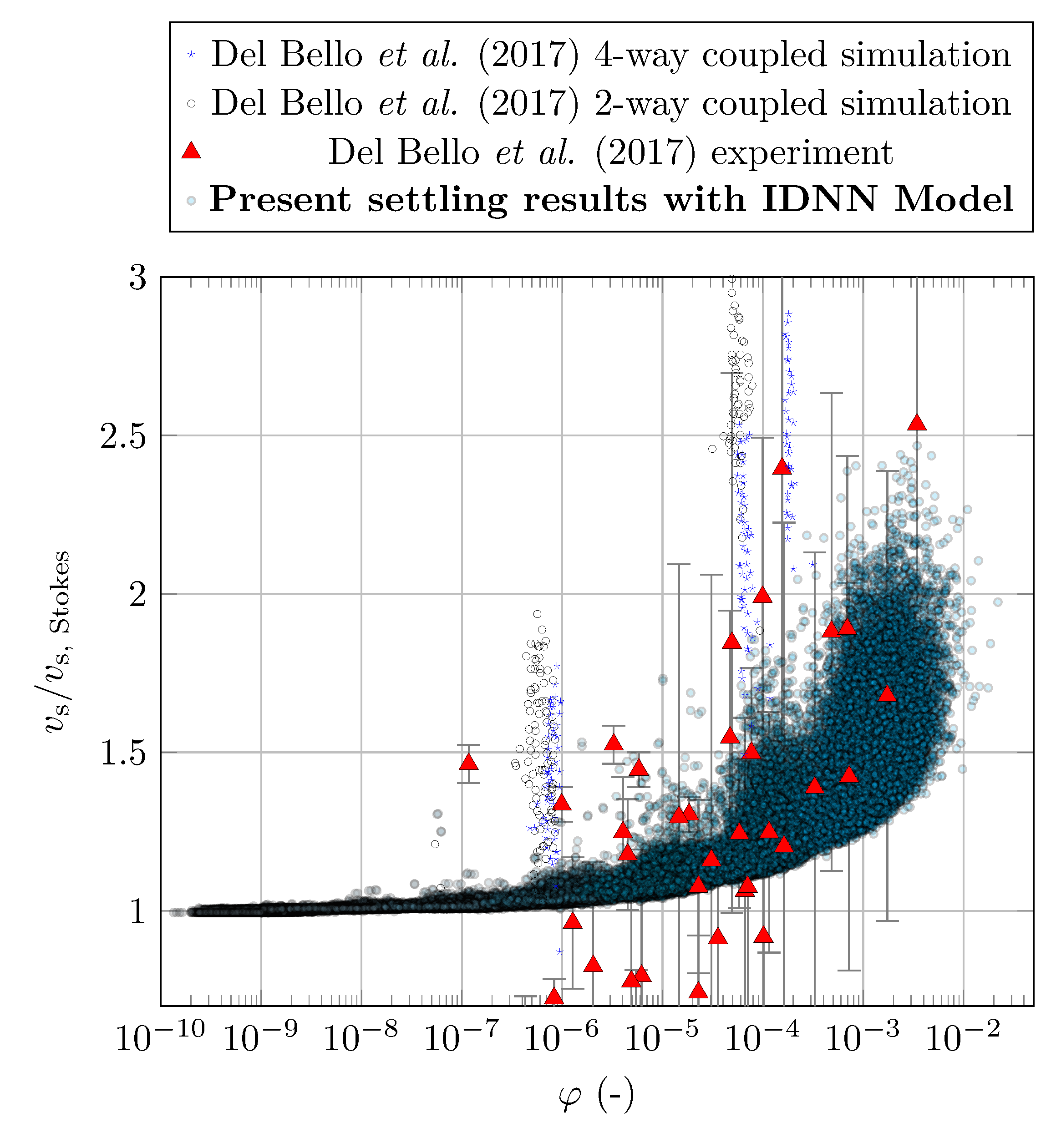}
    \caption{Comparison of the nondimensional settling velocity results of the IDNN model simulation, with the experimental and simulation results of \cite{delbelloEffectParticleVolume2017}, studying sedimenting of the Mt. Etna (Italy) volcanic ash, sized $125$~$\mu$m $< d_\text{p} < 500$~$\mu$m.}
    \label{fig:oneOverBeta}
\end{figure}

The results in Fig. \ref{fig:oneOverBeta} exhibit the behaviour of the increased settling velocity at higher volume fractions. Our results are compared to \citet{delbelloEffectParticleVolume2017}, who studied the sedimentation of the Mt. Etna volcanic ash experimentally, using high-speed camera images and numerically implementing $2$- and $4$-way coupling with fluid. Their results are normalized to the theoretical Stokes settling velocity, using the mode value for the particle size $d_\text{p} = 154.36$~$\mu$m and $\rho_\text{p} = 2600$~kg/m$^3$ for density. In the experiment, the Etna basalt particles demonstrated adequate sphericity and uniformity in both diameter and density, validating their suitability for our simulation, which assumes monodispersed spherical particles \citep{delbelloEffectParticleVolume2017}.

Another way of interpreting the $\beta$ factor is in terms of the modified viscosity, $\eta^* = \beta \eta$ in Eq. (\ref{eq:dragWithBeta}),
\begin{equation}\label{eq:FdStar}
    F_\text{d} = 3 \pi \eta^* d_\text{p} v_\text{s, \text{Stokes}}.
\end{equation}
The observed increase of the settling velocity could be explained by the decrease of the effective viscosity of the fluid. Given our model results, where the reciprocal value of $\beta$ grows with the dispersed phase volume fraction, the modified viscosity $\eta^*$ would then be decreased with the volume fraction, resulting in a decrease of the collective particle drag force, as per Eq. (\ref{eq:FdStar}).

\begin{figure}
    \centering
    \includegraphics{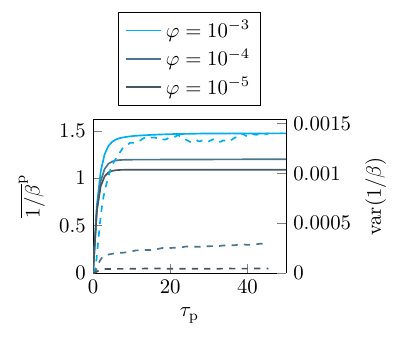}
    \caption{The mean nondimensional settling velocity, $\frac{v_\text{s}}{v_{s, \text{Stokes}}} = 1/\beta$, over time (solid line), and its variance (dashed line). Different colour shades correspond to different bulk volume fractions. The superscript "p", next to the averaging symbol denotes the averaging over the particles.}
    \label{fig:meanVs_varVs}
\end{figure}

To investigate the dynamics of particle motion, we analysed the velocity fluctuations of individual particles using spectral methods. The particle velocity time evolution $v_\text{s} (t)$ was first averaged over the particles to obtain the temporal evolution of mean particle velocity for each bulk volume fraction. Fig. \ref{fig:meanVs_varVs} shows, that after approx. $20$ characteristic timesteps, $\tau_\text{p} = t / \tau$, the system reaches a statistically steady state, as both the mean particle velocity, as well as its variance become stable.

\begin{figure}[h]
    \centering
    \begin{subfigure}{0.4\textwidth}
        \includegraphics{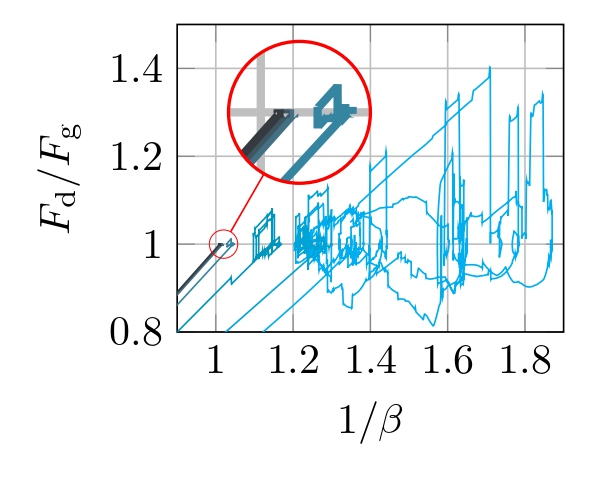}
    \end{subfigure}
    \begin{subfigure}{0.4\textwidth}
        \includegraphics{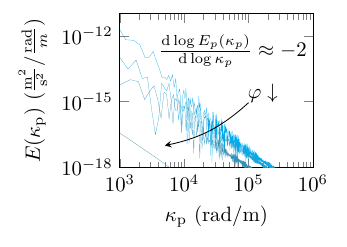}
    \end{subfigure}
    \begin{subfigure}{0.4\textwidth}
        \includegraphics{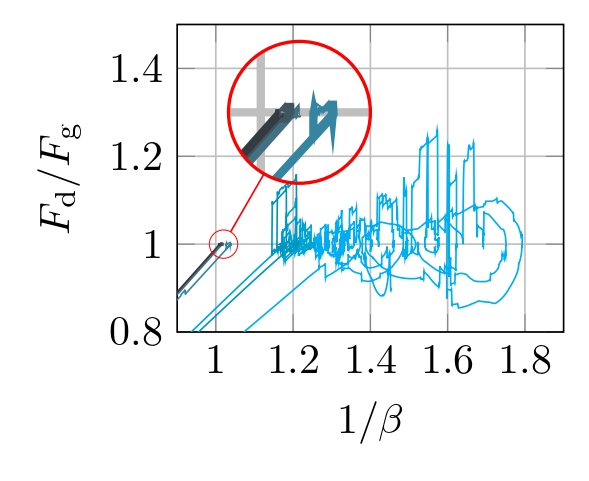}
    \end{subfigure}
    \begin{subfigure}{0.4\textwidth}
        \includegraphics{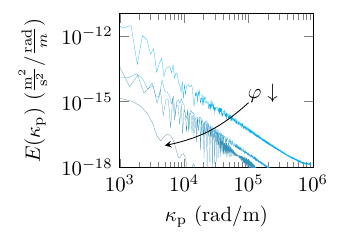}
    \end{subfigure}
    \begin{subfigure}{0.4\textwidth}
        \includegraphics{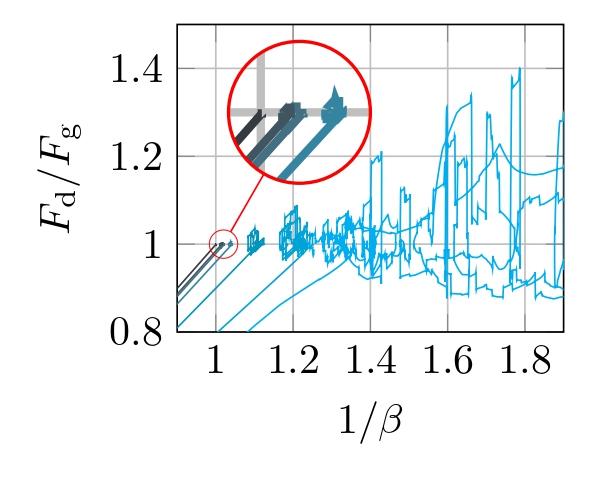}
    \end{subfigure}
    \begin{subfigure}{0.4\textwidth}
        \includegraphics{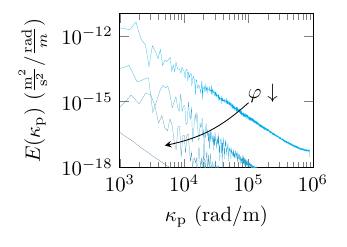}
    \end{subfigure}
    \begin{subfigure}{\textwidth}
        \centering
        \includegraphics[trim = {0cm 4cm 0cm 0cm}, clip]{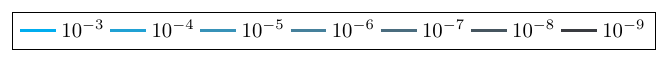}
    \end{subfigure}
    \caption{Phase diagrams (left) showing the dynamics of three randomly selected particles in the sedimenting system, with each row corresponding to one particle. The right panel displays the energy spectra of the same particles' vertical velocity fluctuations, computed via the (FFT).}\label{fig:FFT}
\end{figure}

The dynamic behaviour of individual particles is given in form of a phase diagram on the left panels in Fig. \ref{fig:FFT}. The force on the vertical axes is represented as the actual particle drag force, relative to the gravitational force, $F_\text{g} = |\mathbf{F}_\text{g}|$. Needless to say that in the statistically steady state, the ratio between these forces will equal to unity. On the horizontal axis, the actual particle velocity is presented relative to the theoretical Stokes velocity. We have monitored the dynamic behaviour for three random particles in the system, hence three plots, one on top of the other. The frequency spectra on the right, to which we will return later, correspond to the same particles, presented in the phase diagram on the left. In the dilute limit ($\varphi \rightarrow 0$), each particle settles essentially in isolation. Hydrodynamic interactions vanish, so the drag force exactly balances gravity and the particle velocity approaches the classical Stokes terminal velocity, Eq. (\ref{eq:settlingVelocity}). As the volume fraction increases, hydrodynamic interactions become significant. The phase diagram trajectories no longer collapse into a singular point, but instead scatter around $F_\text{d}/ F_\text{g} = 1$ with $v_\text{s} / v_{\text{s, Stokes}} > 1$. This super-sedimenting behaviour is likely attributed to the local clustering, where, in a real physical system, the downstream particles enter the wake, induced by the leading particle. The downstream particle therefore experiences a reduced drag, briefly accelerating past the isolated sphere velocity. This also likely attributes to the formations of particle clusters, that are known to accelerate past the theoretical limit, and is seen to be more pronounced the higher the $\Ga$ number \citep{uhlmannSedimentationDiluteSuspension2014}. However, as we will show later, the clustering phenomenon is also present to some extent in our case of low $\Ga$ numbers.

The energy spectra on the right side of Fig. \ref{fig:FFT} were created via the fast Fourier transform (FFT) of particle settling velocity over time. To isolate the particle fluctuations, the particle's settling velocity was time averaged, yielding $v_{\text{s},i}' (t) = v_{\text{s},i} (t) - \overline{v}_{\text{s},i}^\text{t}$, where $\overline{v}_{\text{s},i}^\text{t}$ denotes the temporal mean velocity of the $i$-th particle. Note that we use the notation $v_\text{s}$ to denote the settling velocity, as introduced in Eq.~(\ref{eq:dragWithBeta}). This notation emphasizes that we are isolating the vertical component of the particle velocity, even though the IDNN provides the full velocity vector. The superscript $t$ is used to emphasize that the averaging is performed over time, as opposed to averaging across particles, at the statistical steady state, after $50 \tau_\text{p}$. We then computed the power spectral density (PSD) of $v_{\text{s},i}' (t)$ via the FFT, obtaining a two-sided spectrum which was converted to a single-sided PSD, with units of (m$^2$/s$^2$)/Hz. To relate temporal frequencies to spatial scales, we converted the temporal frequency $f_\text{p}$ of the particle velocity to into the spatial wavenumber of a particle $\kappa_\text{p}$ via the relation,
\begin{equation}
    \kappa_\text{p} = \frac{2 \pi f_\text{p}}{\overline{v}_{\text{s},i}^\text{t}}.
\end{equation}
Accordingly, the energy spectrum in the wavenumber space, $K(\kappa_\text{p})$, was computed from the PSD as
\begin{equation}
    E(\kappa_\text{p}) = \mathrm{PSD}(f) \frac{\overline{v}_{\text{s},i}^\text{t}}{2 \pi}.
\end{equation}
The spectrum $E(\kappa_\text{p})$ describes the distribution of kinetic energy among the particle fluctuation magnitudes, caused by the hydrodynamic interactions, analogously to how the turbulent kinetic energy is distributed among eddy sizes. A log--log plot of $E(\kappa_\text{p})$ vs. $\kappa_\text{p}$ was produced to visually examine the resulting spectrum. To quantify the spectral slope, a linear fit was applied to the logarithmic form of the $E(\kappa_\text{p})$ vs. $\kappa_\text{p}$ relation, to find $E (\kappa_\text{p}) \propto \kappa_\text{p}^{-2}$.

The overall shape of the frequency spectrum remains similar across different volume fractions. This indicates that the presence and relative positions of dominant frequencies remains unchanged, but the amplitudes are scaled down. The dominant frequencies are preserved, implying that volume fraction affects only the amplitude of the change of settling velocity rather than introducing new frequencies or shifting existing ones. From the log--log plot is it clear that a power-law decay in spectral amplitude with frequency is observed. This behaviour expresses similarities with turbulence, as the shape of the turbulent power spectra is similar. In the inertial subrange \citet{kolmogorovLocalStructureTurbulence1941} concluded, that the energy scales with $E(\kappa) \sim \kappa^{-5/3}$, where $\kappa$ is the wavenumber of the turbulent oscillations \citep{popeTurbulentFlows2000}. Similarly, we can determine the logarithmic slope for our case, which was computed to be $\sim -2$. If we zoom into each of the plots, we also observe that the intensity of settling velocity fluctuations diminish at small volume fractions.

The pattern observed on the left side of Fig. \ref{fig:FFT} exhibits characteristics of a chaotic system. This behaviour aligns with the fundamental definition of chaos--systems that are theoretically predictable but display apparent randomness over time due to their sensitive dependence on initial conditions. The apparent chaotic behaviour of our particle system is intrinsic to the fact that our multibody behaviour is captured in a series of linear and non-linear transformations while running the IDNN inference during simulation. Let us repeat once more that the force on each particle is computed based on the relative position of its five neighbours. This means that there exists a possibility of those five neighbours either changing order or changing completely during evolution in time.

The fluctuations in particle velocity in our case consist of two contributions: (i) the physical fluctuations, resulting from the multibody hydrodynamic interaction, (ii) the numerical fluctuations, resulting from the fact that the change in the particle drag force is a result of one or more of the five neighbours being changed/switched during the particle cloud evolution. Naturally, the goal of our numerical method is to predict the physical fluctuations in particle velocity, while keeping the frequency of numerical fluctuations as low as possible. In a test, tracking the changes of the five nearest neighbour IDs during sedimentation, it was found that the introduction of entirely new IDs into the neighbour list occurs in less than $1\%$ of the timesteps, with the highest occurrence rate observed at the largest volume fraction, $\varphi = 10^{-3}$, while for lower volume fractions, the number of new IDs was practically negligible. Additionally, the introduction of a new neighbour only happened for the most distant, fifth neighbour.

To further evaluate the behaviour of the particle cloud during sedimenting, clustering was analysed using Voronoï analysis, a widely adopted method for studying particle clustering \citep{uhlmannSedimentationDiluteSuspension2014,oujiaDivergenceConvergenceInertial2020,monchauxPreferentialConcentrationHeavy2010}. To perform the Voronoï analysis, we used the open-source library Voro++, developed by \citet{rycroftVOROThreedimensionalVoronoi2009}. The library takes the list of particle coordinates and performs the Voronoï tesselation. The statistical operations were performed on the resulting Voronoï volumes using python's Scipy toolbox \citep{virtanenSciPy10Fundamental2020}, to obtain the probability density of the Voronoï volumes. 

It was shown by \citet{ferencSizeDistributionPoisson2007}, that for particles that are distributed randomly in a 3D space, the p.d.f of the Voronoï volume closely resembles a gamma distribution. To validate our Voronoï volume calculations, we compared the resulting distribution from an initial random particle configuration at $t=0$ with the gamma distribution (parameters $k=5$, $\theta=0.2$) reported by \citet{oujiaDivergenceConvergenceInertial2020}, and observed a good agreement. Three tests were performed for volume fraction $\varphi=10^{-3}$, as the clustering effect is expected to be more pronounced at higher volume fractions. The $\Ga$ number was then doubled twice, and the Voronoï cell volume p.d.f. was averaged from $50 \tau_\text{p}$ onwards, to obtain the mean p.d.f. in a statistically stationary regime. The obtained p.d.f. widens when transitioning from an initial random distribution to the statistical steady state. This indicates that clusters and voids begin to form at all $\Ga$ numbers once the particle cloud evolves. As the $\Ga$ number increases, the clustering effect is more pronounced, which is observed by the widest p.d.f. being at the highest $\Ga$ number. This observation is in accordance with the DNS study made by \citet{uhlmannSedimentationDiluteSuspension2014}, who also report an increase in clustering with higher $\Ga$ numbers. Since the $\Ga$ numbers in our case are relatively low ($\mathcal{O} (10^{-2}) - \mathcal{O} (10^{-1})$), the clustering is also less prominent.
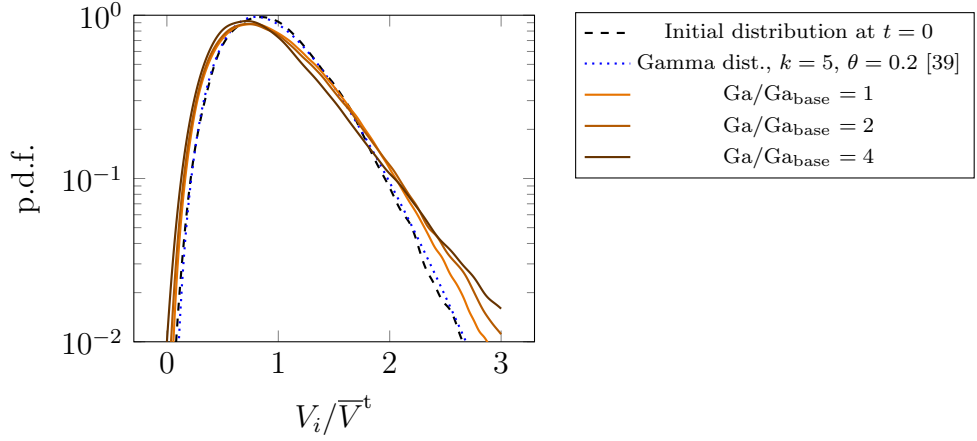
\begin{figure}[h]
    \centering
\begin{tikzpicture}
    \begin{axis}[
        width=0.5\textwidth,
        xlabel={$V_i / \overline{V}^\text{t}$ },
        ylabel={p.d.f.},
        legend pos=north east,
        ymode=log,
        ymax=1,
        ymin=1e-2,
        legend style={
            at={(1.6,0.5)},
            anchor=south,
            legend columns=1,
            font=\scriptsize
        }
    ]
    \addplot[color=black,thick,dashed] table[col sep=comma,x=x,y=kde] {plots/kde_data_zeroth.csv};
    \addlegendentry{Initial distribution at $t=0$}

        \addplot [
        domain=0:3,
        samples=200,
        thick,
        dotted,
        blue
    ] 
    {x < 0 ? 0 : (1 / (24 * 0.2^5)) * x^(5 - 1) * exp(-x / 0.2)};
    \addlegendentry{Gamma dist., $k=5$, $\theta=0.2$ \citep{oujiaDivergenceConvergenceInertial2020}}

    \addplot[color=orange!90!black,thick] table[col sep=comma,x=x,y=kde] {plots/kde_data_Ga.csv};
    \addlegendentry{Ga/Ga$_{\text{base}}$ $= 1$}

    \addplot[color=orange!70!black,thick] table[col sep=comma,x=x,y=kde] {plots/kde_data_2Ga.csv};
    \addlegendentry{Ga/Ga$_{\text{base}}$ $= 2$}

    \addplot[color=orange!40!black,thick] table[col sep=comma,x=x,y=kde] {plots/kde_data_4Ga.csv};
    \addlegendentry{Ga/Ga$_{\text{base}}$ $= 4$}

    \end{axis}
\end{tikzpicture}
\caption{The probability density function (p.d.f.) of the normalized Voronoï cell volumes, where $\overline{V}^\text{t}$ is the time averaged Voronoï cell volume for its respective Ga number case.}\label{fig:voronoiPDF}
\end{figure}

\begin{figure}
\centering
\begin{tikzpicture}
    \begin{axis}[
        width=0.5\textwidth,
        xlabel={$\overline{1/\beta}^{\text{p, t}}$},
        ylabel={$\dfrac{\text{var} (V_i / \overline{V}^\text{t})}{ (k \theta^2)}$},
        enlargelimits=0.1,
    ]
    \addplot[color=black,thick,mark=star] table[col sep=comma,x=vs,y=vars] {plots/phi_vs_varGamma.csv};

    \node[anchor=west, font=\scriptsize, blue] at (axis cs:1.22,1.2381) {$\varphi=10^{-3}$};
    \draw[-stealth, thick,blue] (axis cs:1.33,1.24) -- (axis cs:1.39,1.2381);

    \node[anchor=south west, font=\scriptsize, blue] at (axis cs:1.14,1.1) {$\varphi=10^{-4}$};
    \draw[-stealth, thick, blue] (axis cs:1.175,1.1) -- (axis cs:1.1835,1.04);

    \node[anchor=south, font=\scriptsize, blue] at (axis cs:1.0826,1) {$\varphi=10^{-5}$};
    \draw[-stealth, thick, blue] (axis cs:1.0826,1.0) -- (axis cs:1.0826,0.97);

    \node[anchor=south, font=\scriptsize, blue] at (axis cs:1.02,1.05) {$\varphi < 10^{-6}$};
    \draw[-stealth, thick, blue] (axis cs:1.01,1.05) -- (axis cs:1.01,0.97);




    \end{axis}
\end{tikzpicture}
\caption{Normalized Voronoï cell volume variance, averaged over the statistically steady state, at different volume fractions $\phi$}
\label{fig:varVoronoi}
\end{figure}
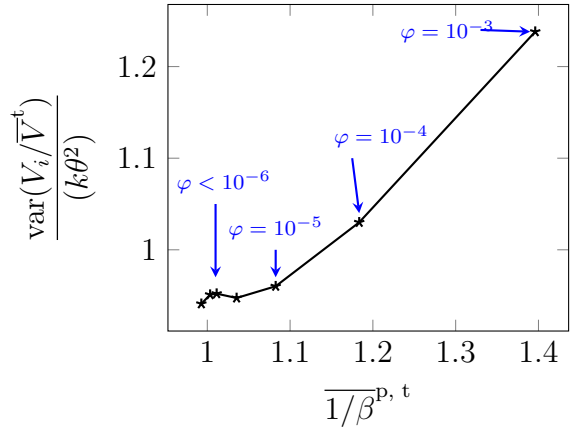

As has been discussed in the past by \citet{uhlmannSedimentationDiluteSuspension2014}, the increased global settling velocity (as seen in Fig. \ref{fig:oneOverBeta}) is caused by the clustering of particles. When clusters are formed, the drafting effect causes a decrease of collective drag force. Although \citet{uhlmannSedimentationDiluteSuspension2014} investigated sedimentation in regimes with higher $\Ga$ numbers, $\mathcal{O}(10^{2})$, where inertial effects dominate and phenomena such as drafting-kissing-tumbling are prominent due to the presence of asymmetric wakes, a similar drafting mechanism may still be anticipated in our regime. Here, the wakes remain axisymmetric and are primarily governed by viscous diffusion, yet hydrodynamic interactions between particles still promote alignment and clustering. This phenomenon is difficult to simulate, as, up to date, DNS was the only approach where the wake behind individual particles could be simulated. Our approach captures these phenomena, as the drafting effect is translated from the BEM database into our simulated sedimenting system. Fig.~\ref{fig:varVoronoi} provides further support for the theoretical framework of particle clustering and its influence on settling velocity within our regime. On the vertical axis we have the ratio between the variance of the Voronoï cell volumes in the statistical steady state and the variance of the gamma distribution from Fig. \ref{fig:voronoiPDF}, which is computed as $k \theta^2$. The variance of the Voronoï cell volume distribution quantifies the degree of spatial inhomogeneity in the particle configuration, with larger variance indicating stronger clustering. On the horizontal axis we have the nondimensional velocity, $1/\beta$, averaged over the particles and over time in the statistical steady state, hence the superscripts "p" and "t" next to the overline. The global settling velocity of particles increases with each increase in bulk volume fraction, accompanied by a corresponding rise in the variance of the Voronoï cell volumes, indicating enhanced clustering.

\begin{figure}
    \centering
    \includegraphics[scale=0.7]{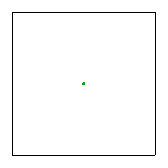}
    \includegraphics[scale=0.35]{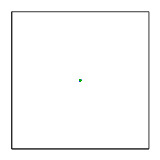}
    \includegraphics[scale=0.35]{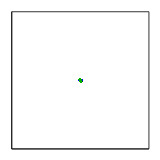}
    \includegraphics[scale=0.35]{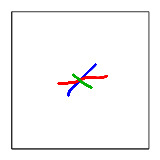}
    \includegraphics[scale=0.35]{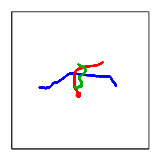}
    \includegraphics[scale=0.35]{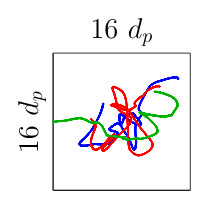}
    \hspace{10em}\includegraphics[scale=0.35]{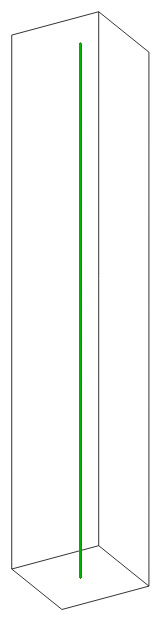}
    \includegraphics[scale=0.35]{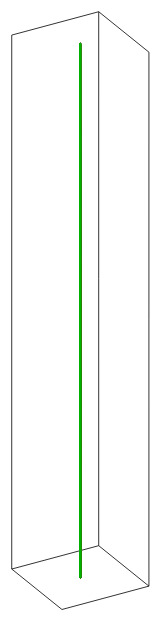}
    \includegraphics[scale=0.35]{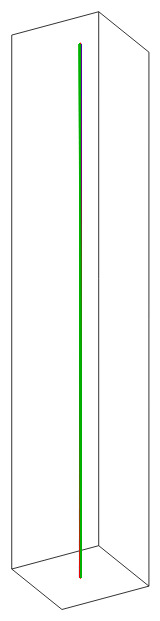}
    \includegraphics[scale=0.35]{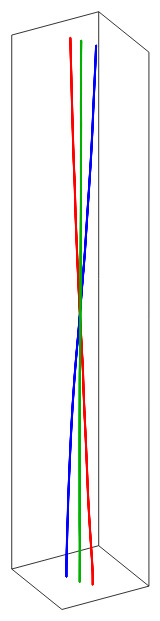}
    \includegraphics[scale=0.35]{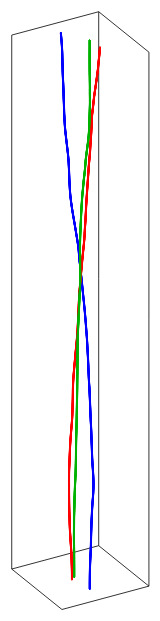}
    \includegraphics[scale=0.35]{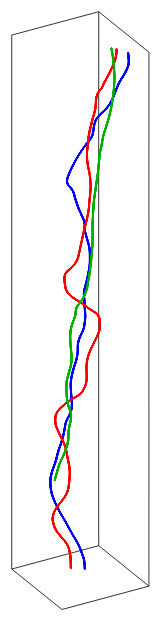}
    \caption{Trajectories of three random sedimenting particles in the computational domain, with each particle shown in a distinct colour (red, green and blue). The bulk particle volume fraction increases from left to right, with values: $\varphi = 10^{-8}$, $10^{-7}$, $10^{-6}$, $10^{-5}$, $10^{-4}$, $10^{-3}$. The top row presents the top view of the system. As described above, the computational domain is cubic with periodic boundary conditions imposed in all directions. Due to these periodic boundaries, the particles undergo multiple vertical passages during sedimentation, and the trajectories are constructed accordingly to reflect this behaviour, hence the elongated vertical dimension of the system.} \label{fig:pathlines}
\end{figure}

The velocity fluctuations are reflected in the instability of the particle trajectories, increasing with bulk particle volume fraction, as seen in Fig. \ref{fig:pathlines}. The trajectory behaviour at high bulk particle volume fractions closely resembles sedimenting at high $\Ga$ numbers. We recall the modified viscosity analogy from Eq. (\ref{eq:FdStar}). Since $\eta^*$ decreases with increasing bulk particle volume fraction, the apparent Galileo number, defined as $\Ga^* = v_\text{s} d_\text{p} \rho_\text{f} / \eta^*$, correspondingly increases.

\section{Discussion and Conclusions}
In this paper, we studied the behaviour of sedimenting solid particles and the effect of their microscopic movement on the cloud as a whole. The numerical approach we have taken was quite different to that taken by previous studies \citep{uhlmannSedimentationDiluteSuspension2014,penlouExperimentalMeasurementEnhanced2023,delbelloEffectParticleVolume2017}, who have either conducted a DNS or a regular $2$- or $4$-way coupled point-particle simulation. By creating a machine learning surrogate, which we named Interaction-Decomposed Neural Network (IDNN), we were able to account for complex interaction phenomena during particle sedimentation, which was previously only achievable by simulating the particle wake formation using DNS. The IDNN works as a black-box, meaning it is integrated into the Lagrangian solver, with its task being to compute a particle drag force vector, based on the relative positions of its five nearest neighbours. Arising from this, a great advantage of the IDNN model is the fact that we are able to obtain the lateral force fluctuations, as well as the fluctuation in the particle streamwise force, by which we physically model the hydrodynamic interaction between particles. By modelling the microscopic particle behaviour, we observe an increase in the collective cloud sedimentation velocity in the dilute regime — consistent with the findings of \citet{delbelloEffectParticleVolume2017,penlouExperimentalMeasurementEnhanced2023} — which we attribute to two effects: (i) the fluctuation in the streamwise particle force around a value, lower than the Stokes limit, (ii) the emergence of particle clusters, sedimenting at higher velocities. Both effects result in an increased mean velocity of the particle, exceeding the Stokes limit. The fluctuations arise from the so-called drafting effect, meaning a persistent entrainment and ejection of particles into and out of the wakes generated by upstream particles. The wakes generated by our particles are axisymmetric and, owing to the low particle Galileo number, are governed by viscous diffusion rather than inertial effects. This results in long-range wake structures that persist over many particle diameters. Relative to the particle size, these diffusive wakes are likely considerably more extended than their inertial counterparts at higher Galileo numbers, which could explain the increase in collective particle velocity deep inside the dilute regime. The energy spectrum of the particles' velocity fluctuations indicate that the hydrodynamic interactions between particles are transferring energy across scales in a manner analogous to turbulent eddies. The large-scale fluctuations impart momentum to nearby particles, generating disturbances at smaller scales, resulting in a decay. At very small volume fractions, the intensity of velocity fluctuations diminishes, leading to a corresponding reduction in the amplitude of the energy spectrum. This indicates that, even at low Galileo numbers, hydrodynamic interactions remain significantly more pronounced at higher volume fractions. Fig.~\ref{fig:meanVs_varVs} demonstrates that the sedimenting system exhibits ergodic behaviour, implying that time averages and ensemble averages converge to the same value. However, this conclusion may be influenced by the persistence of spatial particle clusters. Therefore, further investigation is required to assess the stability and lifetime of these clusters to fully validate the ergodicity of the system.
\clearpage

\section*{Appendix}
\appendix
\section{Particle force computation with BEM}\label{appA}
The governing equation for the steady, incompressible flow of a Newtonian fluid is solved, as described in Eq. (\ref{eq:StokesEquation}). The Stokes flow Green's functions satisfy the continuity equation $\mathbf{\nabla}\cdot\mathbf{u}_\text{f} = 0$ and are the solutions of the singularly forced Stokes equation.
The 3D free-space Green's functions are
\begin{equation}\label{e:StkGF}
	\GF{G}^\star_{ij}=\frac{\delta_{ij}}{r}+\frac{\hat r_i\hat r_j}{r^3}, \qquad
	\GF{T}^\star_{ijk}=-6\frac{\hat r_i\hat r_j\hat r_k}{r^5}.
\end{equation}
The boundary integral representation for the Stokes problem is \citep{pozrikidisIntroductionTheoreticalComputational2011}:
\begin{equation}\label{eq5656}
	c(\gz \ksi) u_j(\gz \ksi) = \int_{\Gamma}^{PV}u_i\GF{T}^\star_{ijk}n_k d\Gamma
	-\frac{1}{\mu}\int_\Gamma \GF{G}^\star_{ji} q_i \text{d}\Gamma,
\end{equation}
where $c(\gz \ksi)=2\alpha$ is twice the solid angle as seen from the point $\gz \ksi$, i.e. in the interior of the domain $c=8\pi$, at a smooth boundary $c=4\pi$. The boundary tractions are denoted by $\mathbf{q} = \gz \sigma\cdot \mathbf{n}$. The normal vector $\mathbf{n}$ points into the domain. The terms on the right represent the double and single layer potentials of the three-dimensional Stokes flow.
To derive a discrete version of (\ref{eq5656}) we consider the boundary $\Gamma = \sum_l\Gamma_l$ to be decomposed into boundary elements $\Gamma_l$:
\begin{equation}
	c(\gz \ksi) u_j(\gz \ksi) = 
	\sum_l\int_{\Gamma_l}^{PV}u_i\GF{T}^\star_{ijk}n_k^{(l)} \text{d}\Gamma
	-\frac{1}{\mu}\sum_l\int_{\Gamma_l} \GF{G}^\star_{ji}q_i \text{d}\Gamma,
\end{equation}
where $n_k^{(l)}$ is the $k$ component of the normal vector pointing from boundary element $l$ into the domain.

Let $\Phi$ be the interpolation functions used to interpolate the function values within boundary elements, i.e. $u_i=\sum_m\Phi_m u_i^{(l,m)}$, where $u_i^{(l,m)}$ is the $m^{th}$ nodal value of function within $l^{th}$ boundary element. Constant interpolation is considered for flux. This yields:
\begin{eqnarray}
	c (\gz \ksi) u_j (\gz \ksi) = 
	\sum_l\sum_mu_i^{(l,m)}\int_{\Gamma_l}^{PV}\Phi_m\GF{T}^\star_{ijk}n_k^{(l)}\text{d}\Gamma 
	-\frac{1}{\mu}\sum_lq_i^{(l)}\int_{\Gamma_l} \GF{G}^\star_{ji} \text{d}\Gamma.
\end{eqnarray}
The following integrals must be calculated for each boundary element $l$:
\begin{eqnarray}\label{eq_int}
	T_{ij}^{(l,m)}(\mathbf{\ksi}) = \int_{\Gamma_l}^{PV} \Phi_m \GF{T}^\star_{ijk}n_k^{(l)}\text{d}\Gamma,
	\nonumber \hspace{0.5cm}
	G_{ij}^{(l)}(\gz \ksi) = \int_{\Gamma_l} \GF{G}^\star_{ij} \text{d}\Gamma.
\end{eqnarray}
Considering boundary conditions we can place the source point into nodes, where unknown values are located and produce a system of linear equations for the velocity and traction. The Andromeda code is able to efficiently simulate Stokes flow based on boundary only discretization. As such it is ideally suitable for performing numerous simulations needed to develop ML based models, as is the subject of present research.
 
Computationally the most expensive part of the simulation is finding the solution of the system of linear equations, created by the BEM based discretization procedure. To facilitate the possibility of parallel computing, we use the  {\it mpich} library  to set up the system of linear equations in parallel and the {\it LIS} library \citep{nishidaExperienceDevelopingOpen2010} to find the solution also in parallel.

\section{Mesh validation study}\label{appC}
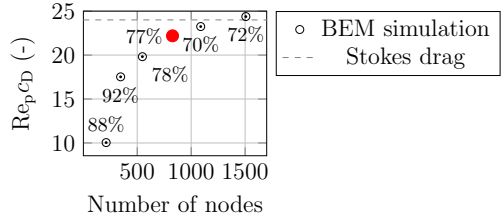
\begin{figure}[h]
    \centering
    \begin{tikzpicture}[scale=0.75]
        \begin{axis}
            [
                ylabel={$\text{Re}_\text{p} c_\text{D}$ (-)},
                xlabel={Number of nodes},
                grid=major,
                width = 0.35\textwidth,
                xmin =1,
                xmax=1700,
                ymax=25,
                legend style={at={(1.05,1)}, anchor=north west},        xticklabel style={
                    /pgf/number format/.cd,
                    1000 sep={}  
                },
                scaled x ticks=false 
            ]

            \addplot[
                only marks, 
                mark=o, 
                color=black
            ] 
            table [col sep=semicolon] {plots/streamwiseForce.csv};
            \addlegendentry{BEM simulation}

            \node[label={\footnotesize $88\%$},circle,fill,inner sep=0.5pt] at (axis cs:212.04141747746138, 10.043483250659897) {};
            \node[label=south:{\footnotesize $92\%$},circle,fill,inner sep=0.5pt] at (axis cs:346.16365514339276, 17.522500350671372) {};
            \node[label=south east:{\footnotesize $78\%$},circle,fill,inner sep=0.5pt] at (axis cs:547.9144616875583, 19.818875046224864) {};
            \node[label=west:{\footnotesize $77\%$},circle,fill,inner sep=0.5pt] at (axis cs:827.1126356460643, 22.189770597161473) {};
            \node[label=south:{\footnotesize $70\%$},circle,fill,inner sep=0.5pt] at (axis cs:1088.3691868249578, 23.245304191479324) {};
            \node[label=south:{\footnotesize $72\%$},circle,fill,inner sep=0.5pt] at (axis cs:1506.7201387383484, 24.380867369709648) {};

            \addplot[
                color=gray,
                style=dashed
            ] table[col sep=space, header=true] {
                X    Y  
                0    24
                2000 24    
            };
            \addlegendentry{Stokes drag}

            \addplot[
                only marks, 
                mark=*,
                mark size=3pt,
                color=red
            ] coordinates {(827.1126356460643, 22.189770597161473)};

        \end{axis}
    \end{tikzpicture}
        \caption{Force exerted by the fluid on a single particle during plug flow versus the number of mesh nodes used. The symbol labels refer to the share of nodes used to discretize the particle, while the rest was used to discretize the outer spherical domain. The mesh chosen for further simulations is shown in red.}
        \label{fig:plugFlowMeshValidation}
\end{figure}

\begin{figure}
    \centering
    \includegraphics{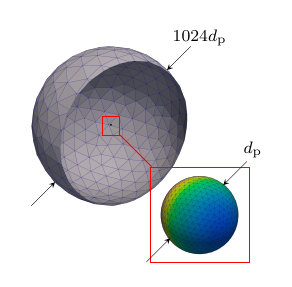}
    \caption{The mesh, recognized as a good compromise between the accuracy and the computational cost, that was used for running numerous simulations during the training database generation. The colour on the particle surface demonstrates the pressure distribution on the particle surface, as a result of the BEM simulation.}\label{fig:finalMesh}
\end{figure}

For this analysis, we focus on the discretization of a single particle in a plug flow and compare the simulated drag force with the analytical solution of the Stokes drag, Eq. (\ref{eq:stokesDrag}). The computational domain is identical to that shown in Fig. \ref{fig:problemDefinition}. A Dirichlet boundary condition is applied to the velocity field on the outer sphere to simulate plug flow, and on the surface of the particle to enforce a no-slip condition. A Neumann boundary condition is imposed on the particle surface for the pressure field. The results plotted in Fig. \ref{fig:plugFlowMeshValidation} show good convergence and the chosen mesh density satisfies both the conditions of good accuracy and computational affordability. For subsequent simulations, where more than one particle is considered in the flow, we keep the mesh design for all particles the same as the particle mesh in the validation study, marked in red. This domain mesh along with the discretized particle, is shown in Fig.~\ref{fig:finalMesh}. We further quantitatively assess the discretization uncertainty by using the method proposed by \citet{celikProcedureEstimationReporting2008}. The BEM numerical method expresses a strong monotone convergence of order $p=2.52$. The numerical uncertainty, in terms of the grid convergence index (GCI), accounts to $8.18\%$. Detailed results are presented in Tab. \ref{tab:GCIresults}. Since the mesh for each of the simulations changes due to changing particle positions, we automized the meshing procedure via Python scripts calling the {\it gmsh} \citep{geuzaineGmsh3DFinite2009} mesher. 
\begin{table}
    \begin{center}
  \def~{\hphantom{0}}
    \begin{tabular}{llr}
        \textbf{Parameter} & \textbf{Symbol} & \textbf{Value} \\[3pt]
        Order of convergence & $p$ & $2.52$ \\
        Coarse mesh extrapolated result & $(\text{Re}_\text{p} c_\text{D})_{\text{ext}, 32}$ & $24.86$ \\
        Fine mesh extrapolated result & $(\text{Re}_\text{p} c_\text{D})_{\text{ext}, 21}$ & $26.32$ \\
        Coarse mesh numerical uncertainty & $\text{GCI}_{\text{coarse}, 32}$ & $2.95\%$ \\
        Fine mesh numerical uncertainty & $\text{GCI}_{\text{fine}, 21}$ & $8.18\%$
    \end{tabular}
    \caption{Results of GCI analysis for plug flow over a single particle.}
    \label{tab:GCIresults}
    \end{center}
\end{table}  

\section{Coordinate system transformations}\label{appB}
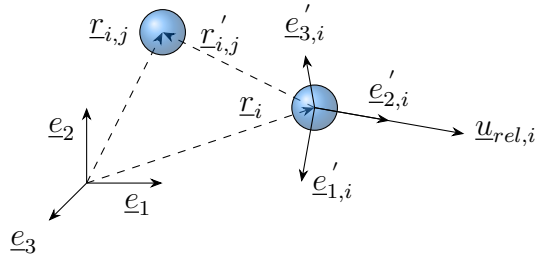
\begin{figure}[h]
    \centering
    \begin{tikzpicture}
        \draw[-Stealth] (0,0) -- (1,0) node[below left] {$\underline{e}_1$};
        \draw[-Stealth] (0,0) -- (0,1) node[below left] {$\underline{e}_2$};
        \draw[-Stealth] (0,0) -- (-135:0.707) node[below left] {$\underline{e}_3$};

        \draw[-Stealth, dashed] (3,1) -- (1,2) node[right, xshift=10pt, yshift=-1pt] {$\underline{r}_{i,j}^{\hspace{2pt} '}$};
        \draw[-Stealth, dashed] (0,0) -- (1,2) node[left, xshift=-8pt] {$\underline{r}_{i,j}$};
        \draw[ball color=cyan!50!blue, opacity=0.5, draw opacity=1] (1,2) circle[radius={0.3cm}];
        \draw[-Stealth, dashed] (0,0) -- (3,1) node[left, xshift=-15pt] {$\underline{r}_{i}$};

        \draw[ball color=cyan!50!blue, opacity=0.5, draw opacity=1] (3,1) circle[radius={0.3cm}];
        \draw[-Stealth] (3,1) -- ++ (-10:2) node[right] {$\underline{u}_{rel, i}$};
        \draw[-Stealth] (3,1) -- ++ (-10:1) node[above] {$\underline{e}_{2,i}^{\hspace{2pt} '}$};
        \draw[-Stealth] (3,1) -- ++ (-100:1) node[right] {$\underline{e}_{1,i}^{\hspace{2pt} '}$};
        \draw[-Stealth] (3,1) -- ++ (100:0.707) node[above] {$\underline{e}_{3,i}^{\hspace{2pt} '}$};        
    \end{tikzpicture}
    \caption{Visualization of the GCS and the LCS.}
    \label{fig:GCS-LCS}
\end{figure}

We observe a cloud of $N$ particles in the fluid flow, where each reference particle, denoted as $i$, is surrounded by a cluster of $M$ closest neighbours, denoted as $j$. The short inter--particle distance causes interactions of the surrounding flow fields, resulting in a disturbance of the reference particle drag force. We consider the cloud of particles in two coordinate system definitions. The global coordinate system (GCS) denotes the global coordinates of the reference particle, 
\begin{equation}
    \underline{r}_i = \underline{e}_1 x_i + \underline{e}_2 y_i + \underline{e}_3 z_i,     
\end{equation}
and its neighbours 
\begin{equation}
    \underline{r}_{i,j} = \underline{e}_1 x_{i,j} + \underline{e}_2 y_{i,j} + \underline{e}_3 z_{i,j}, 
\end{equation}
where $\underline{e}_1 \dots \underline{e}_3$ form the orthonormal base of the GCS and are defined as $\underline{e}_1 = [1, 0, 0]$, $\underline{e}_2 = [0, 1, 0]$ and $\underline{e}_3 = [0, 0, 1]$. The second considered coordinate system is the local coordinate system (LCS) of the reference particle, with the corresponding coordinates denoted as 
\begin{equation}
    \underline{r}^{\hspace{2pt} '}_i = [0, 0, 0],
\end{equation}
\begin{equation}
    \underline{r}^{\hspace{2pt} '}_{i,j} = \underline{e}^{\hspace{2pt} '}_{1, i} x^{\hspace{2pt} '}_{i,j} + \underline{e}^{\hspace{2pt} '}_{2, i} y^{\hspace{2pt} '}_{i,j} + \underline{e}^{\hspace{2pt} '}_{3, i} z^{\hspace{2pt} '}_{i,j},
\end{equation}
where $\underline{e}^{\hspace{2pt} '}_{1, i} \dots \underline{e}^{\hspace{2pt} '}_{3, i}$ form the orthonormal base for the LCS for each reference particle. The subscripts $i,j$ in the above definitions denote that the coordinate corresponds to the $j$--th neighbour of the $i$--th reference particle. The reason behind considering two coordinate systems is that the whole training dataset is defined in the LCS, where the base vector $\underline{e}^{\hspace{2pt} '}_{2, i}$ is aligned with the relative velocity vector at the position of the reference particle, $\underline{e}^{\hspace{2pt} '}_{2, i} || \underline{u}_{rel, i}$, as shown in Fig. \ref{fig:GCS-LCS}.

It can be seen, that in order to obtain the reference particle force in the global coordinate system, a series of transformations has to be applied to the global neighbour particle coordinates. The transformation of the neighbour particle position vector from GCS to LCS can be generally written as
\begin{equation}\label{eq:GCS-LCS}
    \underline{r}^{\hspace{2pt} '}_{i,j} = \underbar{R}_i \left[ \underline{r}_{i,j} - \underline{r}_i \right],
\end{equation}
where $\underbar{R}_i$ is the rotation matrix for the $i$--th reference particle. The rotation matrix has to be constructed, so that the collinearity between the relative velocity vector of the reference particle and the base vector $\underline{e}^{\hspace{2pt} '}_{2, i}$ is satisfied. The rotation matrix can be constructed for two linearly independent vectors, using the Rodrigues' rotation formula \citep{wellerTensorialApproachComputational1998}, which in our case reads as
\begin{equation}
    \underbar{R}_i = 
        c \underbar{I} + \left[ \frac{\underline{u}_{rel, i}}{| \underline{u}_{rel, i} |} \otimes \underline{e}^{\hspace{2pt} '}_{2, i} - 
        \underline{e}^{\hspace{2pt} '}_{2, i} \otimes \frac{\underline{u}_{rel, i}}{| \underline{u}_{rel, i} |} \right] + 
        \left[ 1 - c \right] \frac{ \underline{a} \otimes \underline{a} }{|\underline{a}|^2 },
\end{equation}
where 
\begin{equation}
    c = \underline{e}^{\hspace{2pt} '}_{2, i} \cdot \frac{\underline{u}_{rel, i}}{| \underline{u}_{rel, i} |}
\end{equation}
and
\begin{equation}
    \underline{a} = \underline{e}^{\hspace{2pt} '}_{2, i} \times \frac{\underline{u}_{rel, i}}{| \underline{u}_{rel, i} |}.
\end{equation}
In above equations, the operators $\otimes$, $\cdot$ and $\times$ represent the dyadic product, dot product and the cross product respectively. The above rotation matrix definition holds if $\underline{e}^{\hspace{2pt} '}_{2, i}$ and $\frac{\underline{u}_{rel, i}}{| \underline{u}_{rel, i} |}$ are linearly independent. If the vectors are collinear and contradirectional ($c < 0$), the rotation matrix is constructed as
\begin{equation}
    \underbar{R}_i = - \underbar{I} + 2 \frac{\underline{b} \otimes \underline{b}}{|\underline{b}|},
\end{equation}
where $\underline{b}$ is a vector, perpendicular to $\underline{e}^{\hspace{2pt} '}_{2, i}$. In the case where $\underline{e}^{\hspace{2pt} '}_{2, i}$ and $\frac{\underline{u}_{rel, i}}{| \underline{u}_{rel, i} |}$ are collinear and codirectional ($c > 0$), the rotation matrix is equal to the identity,
\begin{equation}
    \underbar{R}_i = \underbar{I}.
\end{equation}
To be able to use the obtained force prediction in the Lagrangian solver, the obtained prediction must be transformed with the rotation matrix back to the GCS as
\begin{equation}
    \underline{F}_{\text{IDNN}, i} = \underline{R}_i^\top \underline{F}'_{\text{IDNN}, i}.
\end{equation}
\clearpage

\textbf{Acknowledgements.} The authors would like to thank the Slovenian Research and Innovation Agency (research core funding No. P2-0196 and project J7-60118) and the Deutsche Forschungsgemeinschaft (project STE 544/75-1).

\textbf{Declaration of Interests.} The authors report no conflict of interest.

\bibliography{/home/nejcv/Documents/PROJEKTI/bibliography/library}

\end{document}